\documentclass[]{aastex63}
\usepackage{xcolor}
\usepackage{graphicx, natbib, color, bm, amsmath, epsfig, mathrsfs, xfrac, xspace}
\usepackage[inline]{enumitem}
\usepackage{booktabs}
\usepackage{ifthen}
\usepackage{ulem}
\usepackage{acronym}
\newcommand{\myedit}[1]{#1}
\newcommand{\myeditt}[1]{#1}

\acrodef{AIC}[\myedit{AIC}]{\myedit{``accretion induced collapse''}}
\acrodef{DD}[\myedit{DD}]{\myedit{double degenerate}}
\acrodef{DDT}[\myedit{DDT}]{\myedit{deflagration-to-detonation transition}}
\acrodef{HeD}[\myedit{HeD}]{\myedit{helium detonation}}
\acrodef{IR}[\myedit{IR}]{\myedit{infrared}}
\acrodef{LC}[\myedit{LC}]{\myedit{light curve}}
\acrodef{MHD}[\myedit{MHD}]{\myedit{magnetohydrodynamics}}
\acrodef{MIR}[\myedit{MIR}]{\myedit{mid infrared}}
\acrodef{NIR}[\myedit{NIR}]{\myedit{near infrared}}
\acrodef{NLTE}[\myedit{non-LTE}]{\myedit{non-local thermodynamic equilibrium}}
\acrodef{PDD}[\myedit{PDD}]{\myedit{pulsating delayed detonation}}
\acrodef{QNSE}[\myedit{QNSE}]{\myedit{quasi nuclear statistical equilibrium}}
\acrodef{QSE}[\myedit{QSE}]{\myedit{quasi nuclear statistical equilibrium}}
\acrodef{RT}[\myedit{RT}]{\myedit{Rayleigh-Taylor}}
\acrodef{SD}[\myedit{SD}]{\myedit{single degenerate}}
\acrodef{SN}[\myedit{SN}]{\myedit{supernova}}
\acrodef{SNe}[\myedit{SNe}]{\myedit{supernovae}}
\acrodef{SNIa}[\myedit{SN Ia}]{\ac{SN}~\myedit{Type~Ia}}
\acrodef{SNeIa}[\myedit{SNe Ia}]{\ac{SNe}~\myedit{Type~Ia}}
\acrodef{WD}[\myedit{WD}]{\myedit{white dwarf}}

\def\enzo{Enzo}
\def\hydra{\textit{HYDRA}}

\newcommand{\unit}[1]{\ensuremath{\,\, \mathrm{#1}} \xspace}

\newcommand{\myG}{\unit{G}}

\newcommand{\mykm}{\unit{km}}
\newcommand{\mykms}{\unit{km/s}}

\newcommand{\mygvec}{\mathbf{g}}

\newcommand{\myhatbf}[1]{\ensuremath{\hat{\mathbf{#1}}}}

\newcommand{\myrhat}{\widehat{\mathbf{r}}}
\newcommand{\myxhat}{\widehat{\mathbf{x}}}

\newcommand{\myMCh}{\ensuremath{M_{\rm Ch}}}

\newcommand{\myNi}{{\ensuremath{^{56}{\rm{Ni}}}}}
\newcommand{\myCobalt}{{\ensuremath{^{56}{\rm{Co}}}}}

\newcommand{\myfuel}{\rm{fuel}}
\newcommand{\myprod}{\rm{prod}}
\newcommand{\myproduct}{\rm{product}}

\newcommand{\mymathnuc}[1]{\mathcal{#1}}
\newcommand{\myA}[1]{\mymathnuc{A}_{#1}}
\newcommand{\myX}[1]{X_{#1}}
\newcommand{\myY}[1]{Y_{#1}}

\newcommand{\myod}[2]{\frac{ d #1}{ d #2}}
\newcommand{\mypd}[2]{\frac{ \partial #1}{ \partial #2}}

\def\rvec{\ensuremath{\mathbf{r}}}
\def\mvec{\ensuremath{\mathbf{m}}}
\def\Avec{\ensuremath{\mathbf{A}}}
\def\Bvec{\ensuremath{\mathbf{B}}}

\definecolor{orange}{rgb}{1.        ,  0.54,  0}
\definecolor{remove}{rgb}{0.5        ,  0.5,  0.5}

\newcommand{\newtext}[1]{#1}
\newcommand{\newtextb}[1]{{#1}}
\newcommand{\newtextc}[1]{{#1}}

\newcommand\dccq[1]{}

\definecolor{meta}{rgb}{0.371,0.617,0.625} 

\received{today}
\revised{today}
\accepted{today}
\submitjournal{ApJ}

\begin{document}

\title{Physics of Thermonuclear Explosions: Magnetic Field Effects on Deflagration Fronts and Observable Consequences}

\begin{abstract}
We present a study of the influence of magnetic field strength and morphology in
Type Ia Supernovae and their late-time light curves and spectra.  In order to 
both capture self-consistent magnetic field topologies as well evolve our
models to late times, a two stage approach is taken.  
We study the early deflagration
phase ($\sim 1\rm{s}$) using a variety of magnetic field strengths, and find
that the topology of the field is set by the burning, independent of the initial
strength.
We study late time ($\sim 1000$ days) light curves and spectra with a variety of
magnetic field topologies, and infer magnetic field strengths from observed
supernovae.  Lower limits are found to be $10^6\rm{G}$.  This is determined by
the escape, or lack thereof, of positrons that are tied to the magnetic field.
The first stage employs 3d MHD and a local burning approximation, and uses the
code Enzo.
The second stage employs a hybrid approach, with 3D radiation and positron
transport, and spherical
hydrodynamics.  The second stage uses the code \hydra.
In our models, magnetic field amplification remains small during the early deflagration phase.  
Late-time spectra bear the imprint of both magnetic field strength and morphology.
Implications for alternative explosion scenarios are discussed.

\acresetall

\end{abstract}

\author[0000-0001-9556-7576] {Boyan Hristov}
\affil{Center for Space Plasma and Aeronomic Research, University of Alabama in Huntsville, Huntsville, Alabama}
\author[0000-0002-4338-6586]{Peter Hoeflich}
\affil{Department of Physics, Florida State University, Tallahassee, Florida}
\author[0000-0001-6661-2243]{David C. Collins}
\affil{Department of Physics, Florida State University, Tallahassee, Florida}

\section{Introduction}
Thermonuclear \ac{SNe}, or \ac{SNeIa}, are explosions of \acp{WD} \citep{hf60}.
They are important for understanding the Universe and the origin of elements,
and are a powerful tool for measuring large distances. They are also
laboratories for understanding the physics of flames, hydrodynamic
instabilities, radiation transport, non-equilibrium systems, and nuclear and
high energy physics in regimes not accessible by ground-based experiments.
\newtext{ Here,
we examine the impact of magnetic fields on \ac{SNeIa}, which may alter
the sphericity, the nuclear burning front, light curves and spectral properties.}

\newtext{
While \ac{SNeIa} \acp{LC} can be used as 'quasi-standard candles` \citep{p93}, there is growing
observational evidence for spectral
diversity among \ac{SNeIa} that may impact their accuracy as distance measures and, thus, the use of SNe~Ia for precision cosmology.
This has prompted sub-classifications based on observational
characteristics, e.g. high-
and low-velocity \ac{SNeIa} \citep{2005ApJ...623.1011B}, shallow,
core-normal, broad Si-lines, or cool \ac{SNeIa} \citep{2005PASP..117..545B}.
These
classifications are widely used for modern data sets
\citep{2009PASP..121..238B,2013Sci...340..170W,2013ApJ...773...53F}. 
The source of these spectral differences may come
from  similar but
aspherical objects that are seen from different angles
\citep{hoeflich2006b,Motohara06,2010Natur.466...82M,Shen2018}, or
they may indicate a differences in progenitor or explosion scenarios
\citep{hk96,quimby06,shen10,PNK19}. Likely, it is a combination of both.  }

\def\myMch{\ensuremath{M_{\rm{Ch}}}}
\newtext{ A white dwarf, left alone, will eventually cool to the background
temperature of the universe. For the WD to explode, it must interact with a
close companion during the progenitor evolution leading to the explosion. The
companion may either be another WD, (a Double Degenerate System, DD) with short
orbital period, or a non-degenerate star (a Single Degenerate System, SD) such
as  a main-sequence, a helium, or a Red-Giant star.  }

While the exact scenario or scenarios leading to the explosion are still under study, they can be
classified by three distinct mechanisms that trigger the explosion. 
These are; slow compressional heating, surface helium detonations, and the
merger of two white dwarfs.  The first happens on long timescales, while the
latter two occur quickly.
We will discuss each of these in turn.

\newtext{
In the first explosion scenario, the WD accretes material from a companion in either a DD system on long time scales, so called secular mergers, or in a SD system
\citep{WI73,Piersanti2004}.
The explosion is triggered by compressional heat close to the center of the WD
when approaching the \newtextb{critical Chandrasekhar mass,} $M_{Ch}$.
The flame propagates by deflagration
\citep{nomoto84} and, more likely,
starts as a (subsonic) deflagration and transitions to a (supersonic) detonation (\ac{DDT}).
 The  transition from deflagration to detonation is likely 
 due the
 mixing of burned and unburned matter, called \newtextc{the} Zeldovich mechanism.
 \citep{k05,niemeyer96}.\newtextb{ The mixing process can be understood in terms
 of the
 Zeldovich gradient mechanism \citep{Brooker21};} or a unified turbulence-induced mechanism that makes
 \ac{DDT} unavoidable \citep{2019Sci...366.7365P} at densities suggested by observations.  See \citep{2013FrPhy...8..144H} for a detailed discussion.}
 
 \newtext{
 In the second explosion  scenario, 
  a surface
helium detonation (HeD)\footnote{These have also been referred to as DD for  \emph{Double Detonations} in the
literature} triggers a detonation of a sub-\myMCh \ac{WD} with a C/O core.
\citep{wwt80,n82,livne1990,woosley94,hk96,Kromer2010,Sim10,WK2011,Shen2015,Tanikawa2018,Glasner2018,2019arXiv190310960T}.
For triggering the initial detonation, these models require an unmixed surface He layer
of $\approx 10^{-2...-1}M_\odot $ or $\approx 5 \times 10^{-3...-2} M_\odot $
with assumed mixing of carbon into the helium layer on microscopic scales.}

\newtext{
In the third possible scenario, two \acp{WD} merge or collide.   This process
occurs on a dynamical timescale, much faster than the slow accretion timescales
from the previous processes
\citep{webbink84,iben,benz90,rasio94,hk96,segretain97,yoon2007,WMC09,WCMH09,loren09,pakmor10,isern11,pakmor12}.
In simulations of this process, the ejecta show large-scale density asymmetries.
}

\newtext{
 It should also be mentioned that any of these triggering mechanisms can be
 realized within the common envelope of an Asymptotic Giant Branch (AGB) star and a WD.
This has been invoked to explain super luminous SNeIa
 - the so called Super-\myMCh explosions. Here, the degenerate C/O core of the AGB star can explode in two ways. Either, the core grows by accretion on secular time scales from a disrupted WD and compressional
heat triggers the thermonuclear runaway, or the degenerate AGB core merges with
the WD on a dynamical timescale \citep{hk96,yoon2007,soker12,Hoeflich_Book,Hsiao21}.
}

The predominant mechanism for 'normal' SNe~Ia is still under debate  with favorites changing with
time. As we will discuss in the conclusions, nuclear physics dominates the final
outcome of the explosion, and successful models result in almost overlapping
progenitor mass ranges among different scenarios.

\newtext{In our study we examine the impact of magnetic fields on several
aspects of \ac{SNeIa} in  the classical delayed-detonation scenario because this
allows \newtextc{us} to reproduce light curves and spectra of classical SNe~Ia.} The results
will be put into context of other explosion scenarios in the final discussion
and conclusions. \dccq{Do we want to say anything stronger about why DDT, or
leave it?}

\newtext{ Our study is based on and combines two previous papers.
\citet{Hristov2018} (Paper I) showed the magnetic field effects on nuclear
burning fronts
in an rectangular tube. \citet{penney14} (Paper II) studied the 
effect of magnetic fields for dipole and turbulent morphologies on positron
transport, which showed the presence of high magnetic fields
($B$-fields) in some observed supernovae. The goal of this paper is to extend and combine these
studies for full star simulations with initial conditions which allow to
reproduce observations, and to test the basic assumption on the field morphology
in Paper II.}

The obvious question concerns the origin of high $B$-fields. While some WDs are
observed with $B$ of several $10^7\myG$, the majority have no measurable field
\citep{liebert03,schmidt03,Silvestri07,tout08}. As will be discussed in the
conclusion, high $B$ would require field amplification: a) by rotation induced
circulation b) during the  smoldering phase, which is characterized by convection driven non-explosive carbon burning, with large Eddie sizes corresponding to the pressure scale
height in the WD \citep{hs02}; or c) during the dynamical phase of the explosion considered here.  

For the \ac{DDT} mechanism, one of the crucial problems is how to partially suppress the strong \ac{RT}  instabilities during the explosion
\citep{1995ApJ...449..695K,1997NewA....2..239N,gamezo03,roepke05,hoeflich06,Motohara06,penney14,tiara15,lluis2019,Yang2019}.
\newtext{
This instability forms when low-density, high-temperature material is accelerated into higher density
material. The acceleration is most commonly a gravitational field, but can be also bulk acceleration.
The instability manifests as a pattern of rising mushroom-like structures of low density burned material,
leading to large scale  mixing.
In particular, 3D hydrodynamical simulations predict rising plumes throughout the entire WD \citep{gamezo03,roepke05}. Although
plumes \newtextc{at} the RT scales have been observed in supernovae remnants \citep{fesen07} and indicated by high-resolution polarization spectra of SNe~Ia \citep{patat12,Yang2019}, they are very constrained in velocity space.
High magnetic fields are known to suppress the RT instability \citep{chandra61book},
and as a WD is a fully ionized plasma, it is reasonable to consider the effect
of magnetic fields \citep{Remming14,Hristov2018}}.

In Paper I, we simulated a 
  $240\rm{km}\times 15\rm{km}\times 15\rm{km}$ tube of constant density with WD conditions with magnetic fields from $10^9\rm{G}$ to $10^{12}\rm{G}$.  
Our finding was that the development of \ac{RT} instability was partially suppressed at magnetic field strength of $10^{10}\rm{G}$, and almost
completely suppressed at $10^{12}\rm{G}$.  
 In most cases, this was related to a reduction in the overall
burning rate.  However, certain configurations 
of burning front and magnetic field (namely models Z12 and YZ12 where the field
 is strong and aligned with the propagation direction) 
  the front speed and thickness of the burning region are increased
relative to
the unmagnetized case.  In this configuration, the field maintains a thicker
burning region, and the increased burning increases the speed of the front.
If this configuration manifests in a real star, it may give another route to
the DDT. 

\newtext{Paper II studied the effect of magnetic fields for dipole and turbulent morphologies on the 
positron transport, and found that fields can greatly impact
the late-time light curves and line profiles.   The size of the effect greatly depends on the morphology of the field.  In previous studies, the magnetic fields have been assumed (a) to be radial or locally-trapping small-scale fields \citep{Milne2001}; or (b) to be dipoles or arbitrary turbulent fields \citep{Hoefich04}}. 

\newtext{The trapping of positrons by magnetic fields can be observed as an increase in
brightness in the late-time light curve, and in the width and evolution of certain atomic line profiles.
Between 300 and 1000 days after the explosion, the primary energy source in the
supernova is positrons produced by the decay of \myCobalt.  If the positrons are trapped by the magnetic field, the supernova will remain brighter than if they are allowed to freely diffuse away from their birthplace.}
 The degree of
trapping depends on both the morphology and strength of the magnetic field, and
has been subject to several PhD theses \citep{Milne1999,penney12}. 
For light curve studies, \citet{Milne2001} calculated positron transport for
radial magnetic fields and assumed that, for turbulent  fields, all positrons are
locally trapped. 

In Paper II, we used small scale turbulent and dipole fields as representation of large fields. 
We showed that positrons are not locally trapped even for turbulent fields. 
The results depend on the size, scale and morphology of the field. We also showed that
magnetic fields and subsequent positron trapping can also impact late-time
\ac{NIR} spectra. Specifically we focused on the 1.644 $\mu m$ feature, which is
dominated by a single [FeII] transition, rather than consisting of other blended
features in the optical and \ac{NIR} \citep{h04,tiara15}.  
After about day 300, the shape of the line profile
provides important information about the distribution of \myNi, including asymmetries, which may in part come from magnetic fields
\citep{h04,Motohara06,tiara15,2018MNRAS.477.3567M}. 
\newtext{Using this line, lower limits for the magnetic field in SN2003du field
were found to be $5000 G$ using analytical approximations for the positron
transport \citep{h04}.  } Detailed positron transport calculations and
observations of \ac{NIR} line profiles suggest lower limits larger by one to two
orders of magnitude. 

{ The 1.644 $\mu \rm{m}$ line also opened a new aspect and further
motivates our focus on DDT models. While the line offset may be influenced by peculiar velocity of the progenitor system relative to the host galaxy (typically 100 \mykms), it also reflects the orbital velocity of the binary system.}
Narrow binary systems result in large
offsets in velocity, whereas wide binary systems (and lower velocity
offsets) are expected when the companion star is
non-degenerate.   Though the statistics is small, the offset shift ($v < 500 ...
1,000 km/sec)$  of the feature in several \ac{SNeIa} (
SN2003du \citep{h04},
SN2005df \citep{diamond2015},
SN2014J \citep{Diamond2018},
SN2012ht,
SN2013aa, and
SN2012cg \citep{2018MNRAS.477.3567M}) suggests the existence of wide progenitor
systems rather than narrow systems to be expected for single-degenerate
progenitors.  \newtextc{Two possible exception are SN2012fr and SN2013cf} \citep{2018MNRAS.477.3567M}, which show
velocities consistent with double-degenerate progenitors \citep{Shen2018}.  Both narrow and wide systems are can lead to \myMCh\ explosions, whereas mergers and helium detonation models can only originate from close binaries with large velocities.

This work extends and connects two prior studies with a focus on the morphology and size of the magnetic field. Namely, it tries to answer the following questions: Do the conclusions from simulations in the box still hold up for full-stars with density gradients?
Does a deflagration phase change the morphology of an initial dipole field? Can the high magnetic fields be produced during the early deflagration phase?
What constraints can be put on the size of the magnetic fields in the progenitor without arbitrarily assuming its morphology? 

\newtext{The goal of this paper is to study the effect of magnetic fields on the
explosion and the observational consequences. For computational feasibility, we
do not present a single end-to end simulations for the entire evolution but
strive to
perform two interconnected and consistent sets of simulations.
In the first part, we assume the size of the magnetic field and
determine its morphology from simulation. In the second part, we assume the morphology
of the magnetic field and constrain its size from light curves.  It is important to address the
question whether observations can constrain the magnetic field into the regime
where they cannot be neglected.  }

{
Here, we run two stages of simulations. For consistency, we start with a WD
structure which allows \newtextc{us} to reproduce observations, and consistent equations of
state throughout all phases.  The first stage extends the study of Paper I to a
three dimensional star to include large-scale wave modes.  The ability of the
magnetic field to suppress instability is a function of the wavelength of the
perturbation.  In Paper I, we used a small rectangular domain, which restricted
perturbations to those that can fit across the box, 15\rm{km}.  While we were
able to completely suppress RT, larger wavelengths may still be unstable.  This
stage uses 3D
magnetohydrodynamics (MHD), with the size of  $B$ as a free parameter, and a
simplified burning model, starting from a stage at which Rayleigh-Taylor
instabilities are known to develop. The second stage extends Paper II by
employing the magnetic field topology from a more realistic model, namely the
simulations of the first stage. In the second stage we study the impact of
magnetic field strength and topology on positron escape in the late stages (1000
days) of light curves and line profiles to constrain the size of the
initial $B$-field.  This stage uses 1D radiation hydrodynamics, a more complete
nuclear network of 218 isotopes, time-dependent non-LTE models for atomic level
populations,  and 3D photon and positron transport.  We assume that the magnetic
topology is set by the deflagration phase, and that the post-detonation dynamics
are entirely determined by the explosion.  We then simulate the light curve as
described in \citet{Hoeflich2017}, with the addition of a three dimensional
magnetic field that is passively advected along with the (homologous) expansion
of the supernova.  This allows us to more accurately capture the behavior of the
positrons, while remaining computationally feasible. 
}
 
Our paper is organized as follows. 
In Section \ref{method} we will describe the numerical methods employed,
  and the setup of the simulations. 
  In section \ref{sec.evolution}, we will 
  discuss stage 1, results of the early deflagration and the morphology of the field.
  In section \ref{sec.stage2},  we present stage 2, and will address the impact of magnetic fields on late-time spectra and
  light curves.
  In section \ref{sec.discussion}, we will discuss the results and
  implications for the scenario-dependent observables. We
  summarize our results, discuss the possible sources of amplification, and
  briefly
  address the differences between \myMCh ~and helium-triggered explosions.

\begin{figure}[h]
\includegraphics[width=.5\textwidth]{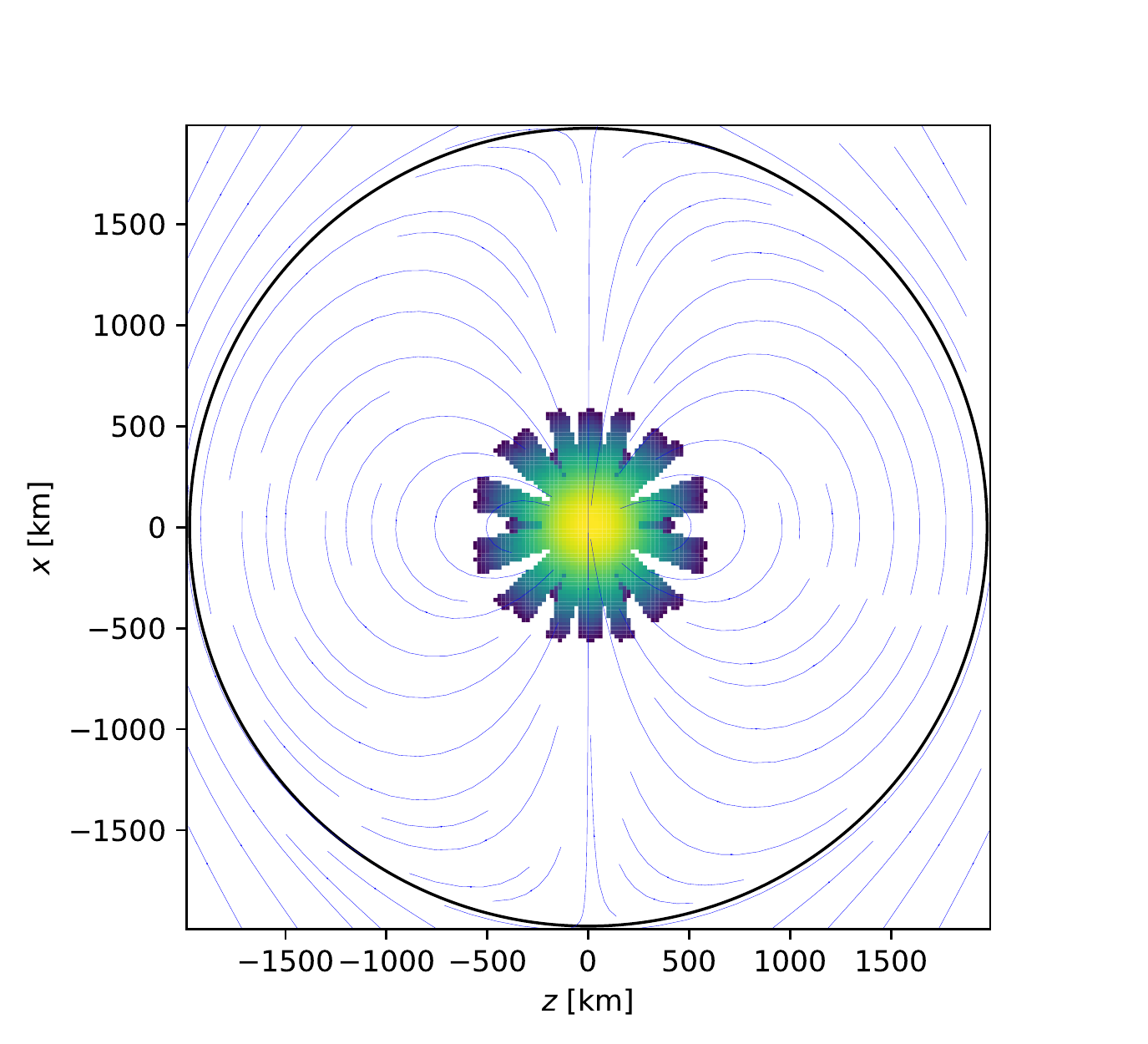}
\includegraphics[width=.5\textwidth]{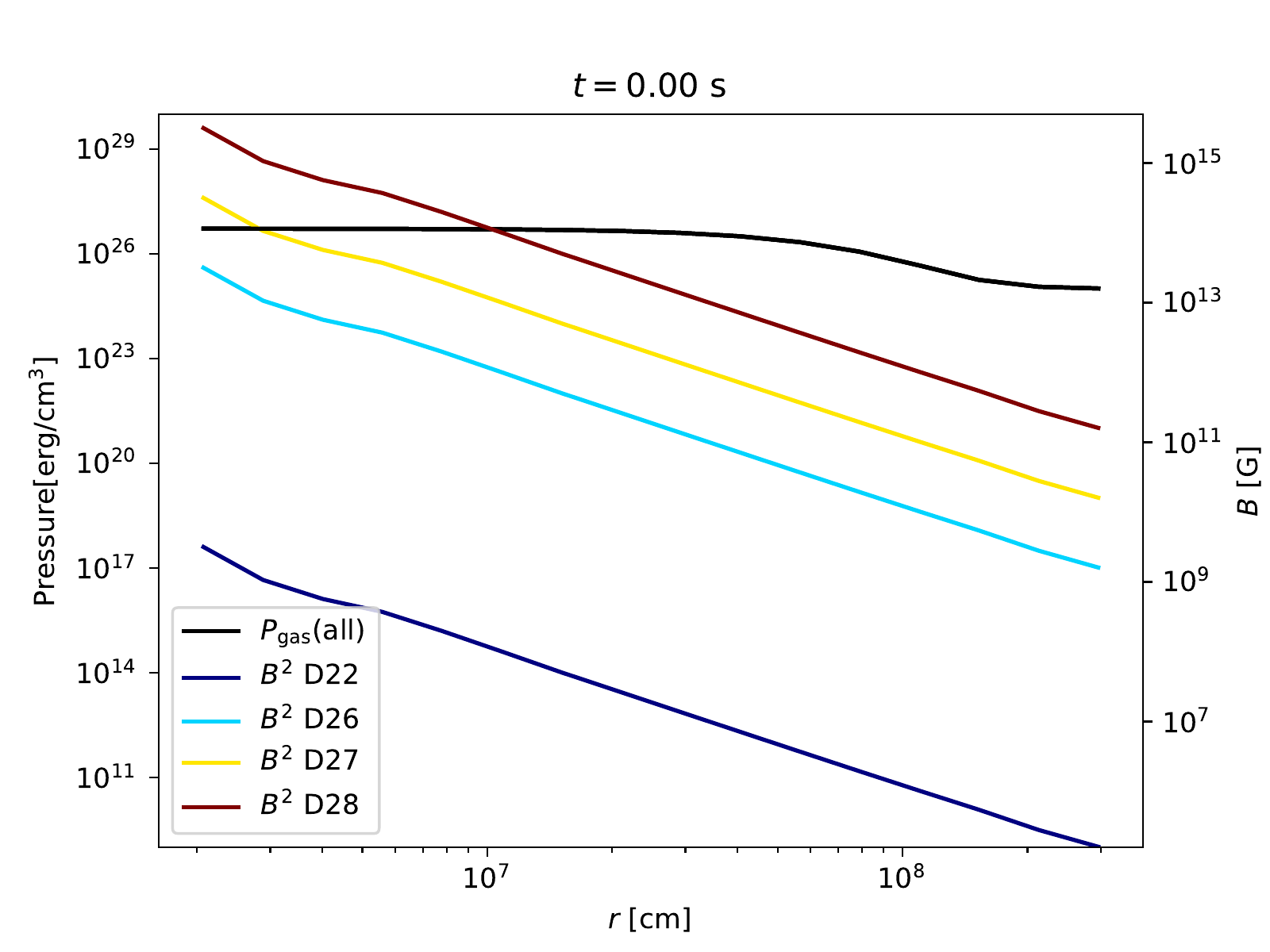}
\caption{
{\bf
Initial Configuration. } \emph{(left)} A projection of the initially burned
region and magnetic field lines.  The black line shows the edge of the star at
1975 km.  The initially burned material is distributed in fingers radiating from
the center.  A total of 80 fingers starting 200 km from the center
\newtextb{with} length 400 km are used. This initial
configuration is identical for all 4 simulations, with only the strength of the
field changing.  \emph{(right)}
The initial thermal pressure (black line) and magnetic pressures vs. radius for
all four simulations. The black line shows the pressure profile (identical for all
runs) while the colored power law lines show the magnetic pressure (left axis) or
field strength (right axis).
}
\label{fig:prisms}
\end{figure}

\begin{deluxetable}{cc}
\tablecaption{Model (run) names vs. the strength of the
initial magnetic field, $B$.
\label{tab:runmatrix}}
\tablehead{
\colhead{Model Name} & \colhead{Magnitude of the Initial} \\ 
& \colhead{Magnetic Dipole Moment}\\
& \colhead{[$\rm{G\ cm^{-2}}$]}
}
\startdata
D22 & $ 10^{22} \hat{x}$ \\ 
D26 & $ 10^{26} \hat{x}$ \\ 
D27 & $ 10^{27} \hat{x}$ \\ 
D28 & $ 10^{28} \hat{x}$ 
\enddata
\end{deluxetable}

\section{Method}
\label{method}
   
Our two stages of simulations both begin from the same hydrostatic white dwarf.
The initial model for all simulations in both stages is based on Model 23 from
\citet{Hoeflich2017}.

The first stage is restricted to the early  deflagration
phase; this phase explores the impact of magnetic fields on the burning, and examines the evolution of the topology of the magnetic field.
This stage follows the non-distributed regime of burning in three
dimensions with \myeditt{3D} \ac{MHD}.

The second stage also begins from Model 23, and follows the SNIa to late times, examining the impact of magnetic field strength and topology on the transport of positrons, and the resulting effect on light curves and spectra.
Stage 2 uses a hybrid approach, employing distributed burning in 1D,
while the 3D magnetic field is a passively carried with the explosion.
To initiate the magnetic field in the second stage we map the magnetic 
field from stage 1 onto a spherical comoving grid.
We follow the evolution to the phase of free expansion.
This two stage approach allows us to 
use full \ac{MHD}
when the magnetic field is thought to be
dynamically important and to create self-consistent magnetic topologies, 
but also follow the light curve for much longer than
possible for an explicit 3D \ac{MHD} simulation.

The first stage, discussed in Section \ref{enzo}, uses the 3D \ac{MHD} code \enzo\ \citep{Bryan14}. 
For the second stage, discussed in Section \ref{hydra}, we use our
\ac{NLTE} code for 3D HYDrodynamcal RAdiation transport \hydra.  
Results are discussed in Sections  
\ref{sec.evolution} and
\ref{sec.stage2}, respectively.

\subsection{Stage1: Initial Deflagration}
\subsubsection{White dwarf setup}
\label{wdsetup}

Each of our four stage 1 simulations begins with a hydrostatic WD with a central density of 
$1.15 \times 10^9 \rm{g/cm^3}$ and an initial radius of 1975 km.  
To seed the burning instability,
we begin with the central region initially burned in a star-like
pattern.  
We superimpose a magnetic dipole oriented along $\myxhat$ at the origin.
The only parameter that changes between the four models is the dipole moment magnitude, i.e. the strength of the magnetic field.
The cartesian domain is $4000 \mykm\mathbf{^3}$, with outflow boundaries on all quantities. 

Model parameters are chosen to resemble the physical conditions of a WD after
the onset of the deflagration stage, while the burning is still in the
non-distributed regime. 
We interpolate from Model 23 to the higher resolution 3D grid using
 Taylor series for the density and the pressure, subject to the
requirement to fit the original profiles and derivatives at the innermost
points.

The internal energy is initialized to maintain hydrostatic equilibrium in the
unmagnetized case.
The value of the adiabatic index, $\gamma = 1.35$ was chosen
to be consistent with stage 2.
This choice is baesd on the equation of state \newtextb{(EOS)}
as implemented in HYDRA (see appendix B, and Paper I).
The result is a radial pressure profile,
not including magnetic pressure, from which we construct the internal energy
profile.  
Owing to the combination of the mismatch between the spherical structure and
Cartesian grid and the magnetic pressure, we were left with small residual
velocities within the inner 100 km, which we discuss further at the end of this
section.

Most of the star is pure fuel (50:50 carbon/oxygen).
We initialize a burning front at the center (100\% $^{56}\rm{Ni}$).
To build the burning front, we first triangulate a sphere with radius 200 km with 80 facets, then add 200 km fingers on each
facet.  A projection of this star-like pattern can be seen in Figure
\ref{fig:prisms}.

The initial magnetic field is a global magnetic dipole with a moment along the $x$-axis.
To ensure the divergence of the magnetic field is numerically zero, we initialize the 
vector potential and take its curl to produce the magnetic field.  The vector
potential is
\begin{align}
    \Avec(\rvec) = \frac{\mvec \times \rvec}{4 \pi r^2},
    \label{eqn.dipole}
\end{align}
where $\mvec=M \hat{x}$.  The initial dipole moment $M$  are  $10^{22}, 10^{26},
10^{27}, \text{and } 10^{28} \, \rm{G}{cm^{-2}}$.  We refer to each simulation by the order of magnitude of the
dipole moment, that is $D22, D26, D27,$ and $D28$, respectively, see Table
\ref{tab:runmatrix}.  The initial magnetic topology can be seen in the left
panel of Figure \ref{fig:prisms}.  The angle-averaged magnetic profile can be
seen in the right panel of that figure, which shows the magnetic pressure and
gas pressure (left axis) and magnetic field strength (right axis).  

There are two extraneous sources of acceleration in our setup: the first from
errors in mapping the spherical star to the Cartesian grid; and the second from
our non-force-free magnetic configuration.  These accelerations 
are confined to the inner 100 km, which is well within the initially burned
region, and only imparts a small amount of excess kinetic energy.  It should
be noted that simply offsetting the magnetic pressure by reducing the gas
pressure is insufficient, as the magnetic tension
$\mathbf{B}\cdot\nabla\mathbf{B}$ must also be initially zero. This must be
done numerically.
While force-free magnetic fields should be used in future experiments, the perturbation
here is not large enough to affect our conclusions.

\subsubsection{Stage 1 method: \enzo}
\label{enzo}

For the first stage of simulations, we use the code \enzo\ \citep{Collins10, Bryan14}
modified for non-distributed burning.
The code is designed for astrophysical applications supporting
a number of physics processes, including \ac{MHD}.
It solves the Eulerian equations of ideal \ac{MHD} 
as well as the non-distributed burning equations,
which are:

\begin{equation} \label{eq:MHD1}
\mypd{\rho}{t}
+ \nabla \cdot
(\rho \mathbf{v})
= 0
\end{equation}

\begin{equation} \label{eq:MHD2}
\mypd{\rho \mathbf{v} }{ t}
+ \nabla \cdot
\left( 
	\rho \mathbf{v} \mathbf{v} 
	+ \mathbf{I} P 
	- \frac { \mathbf{B} \mathbf{B} }{ 8 \pi } 
\right)
=
- \rho \mathbf{g}
\end{equation}

\begin{equation} \label{eq:MHD3}
\mypd{E}{t}
+ \nabla \cdot
\left[
	(E + P) \mathbf{v} 
	-
	\frac 
	{ \mathbf{B} ( \mathbf{B} \cdot \mathbf{v} ) }
	{ 4 \pi }	
\right]
=
- \rho \mathbf{v} \cdot \mathbf{g} 
+ \dot{Q}
\end{equation}

\begin{equation} \label{eq:MHD4}
\mypd { \mathbf{B} }{ t}
+ \nabla \times ( \mathbf{v} \times \mathbf{B})
=0.
\end{equation}
Here, $\mathbf{v} \mathbf{v}$ and $\mathbf{B} \mathbf{B}$ are 
the velocity and the magnetic field outer products,
$\rho$, $\mygvec$, and $\dot{Q}$ are the density, 
the gravitational acceleration, 
and the rate of energy production from the nuclear burning.
Further 
$E = e + { \rho v^2 } / { 2 } + {B^2} / {8\pi}$
is the total energy density,
$P = p + {B^2} / {8\pi}$
is the total pressure.

The following equation of state closes the system:

\begin{equation} \label{eq:EOS}
e = \frac{p}{\gamma - 1}
\end{equation}

To speed the calculation and reduce numerical instability, the gravitational acceleration, $\mathbf{g}$, is computed from the spherically
averaged density,
\begin{align}
    \mathbf{g}(\mathbf{r}) = \myrhat \frac{4 \pi G}{r^2}\int_0^r \rho r^2
    dr.\label{eqn.gravity}
\end{align}
\newtext{To check the validity of this assumption, we tested this solver against the FFT-based gravity solver in
Enzo, using the strongly magnetized (and least round) case.  Deviations between the two were at
most one percent at the outer boundary, and $<<1\%$ in the interior.}

\newtext{In the non-distributed burning approximation, we assume that the width
of the burning front
is small compared to a zone, and that the fuel is burned instantaneously to
product and specific energy, Q. 
Our burning operator is based on \cite{k05}. In this model, the flame propagates via diffusion 
of the molar burned fraction, $f$
}
\begin{equation} \label{eq:flame-diffusion}
\mypd{f}{t} + \mathbf{v} \cdot \nabla f
= K \nabla^2 f + R,
\end{equation}
where $f=0$ is pure fuel.  $K$ and $R$ are the diffusion rate and reaction rate,
respectively. \newtext{ $R=R_0=const$ if $f$ is between the threshold for burning, $f_0=0.3$, and unity,
when no burning is possible, and zero otherwise.  The constants are chosen such
that the front diffusion speed $D_f = \sqrt{KR_0/f_0} = 100 \mykms$.
The energy produced by the nuclear burning, $\dot{Q}$, needed in \ref{eq:MHD3}, is calculated as:}
\begin{equation} 
\label{eq:Qdot}
\dot{Q} = \mathcal{Q}_{\rm{burn}} 
\myod{\rho_{\rm{prod}}}{t}
\end{equation}
\newtext{where $\mathcal{Q}_{\rm{burn}}$ is the nuclear  energy released per gram fuel and  
$\rho_{\rm{prod}}$ is the partial mass density of the burned product.}
The relation between $f$ and $\rho_{\rm{burn}}$ can be seen in Equation
\ref{eqn.frac_prod} in Appendix \ref{appx}.

We use the Constrained Transport \ac{MHD} module in \enzo
\citep{Collins10,Bryan14}.  We have \newtextc{shown} that this method preserves $\nabla\cdot \Bvec=0$ to near machine precision. 
For the primary evolution of the MHD equations, we use \citet{Lilice08},
and the HLLD Riemann solver of \citet{Mignone07}. To compute the electric field, the Constrained
Transport method of \citet{GS05} is used.  

\subsection{Stage 2: Late time evolution}
\label{stage2}
\newtext{We wish to study the impact of magnetic field strength and topology on the
transport of positrons in late-time ($\sim 1000$ day) supernovae, and their
impact on light curves and spectra.
We show that there is a measurable increase in late-time light curves ($\sim
0.5^m$) if the positrons are locally trapped by the magnetic field, rather than
the unmagnetized case which allows for positrons to more freely diffuse out of
the core.  This can be used to estimate lower limits on magnetic field strengths
of published supernova, see Section \ref{sec.stage2}.  }

\subsubsection{Remapping and hybrid model}
\label{stage2remap}

\newtext{The cost of continuing
the \enzo\ simulations until t=1000 days is computationally prohibitive; one of these
simulations took several days on 64 processors to compute one second of time.  As there are
$8\times 10^7$ seconds in 1000 days, continuing the run would surely extend past the end of the
funding period. Moreover, the physics needed to follow the explosion beyond the phase of non-distributed burning
are not currently available in \enzo.  Much success has been had by employing 1D
simulations, which can afford more complete
physics packages \citep{Hoeflich2017}.  
However, the diffusion of positrons is a
fundamentally three dimensional process, being dependent on the strength,
topology, and distribution of the magnetic field.  
Thus, for stage 2, we adopt a hybrid approach, where the hydrodynamics is
solved in 1D (assuming spherical symmetry), while the positron transport is 3D.
The 3D magnetic field is taken as a passive
tracer and expands in a frozen-in manner with the expanding supernova.}

The core assumption is that the magnetic field topology
 is determined during the subsonic deflagration phase, and the dynamics during the supersonic
 detonation phase are determined exclusively by the overwhelming explosion energy.
 We show in Section \ref{sec.evolution} that only the strongest magnetic field has any
 impact on the burning front; the three more reasonable cases are quite
 spherical and have similar topologies, which justifies our neglect of the
 back-reaction of the magnetic field on the burning in Stage 2.  While it is possible that the topology may continue to evolve after the detonation, it is unlikely that the field will become less tangled, and it is unlikely that the field will be further amplified.  As the decay timescale for a magnetic field in a fully ionized WD with temperature $T=10^9$K is $10^9$ yr \citep[see, e.g.,][]{choudhuri}, it is unlikely that substantial decay will take place in the few hundred days examined here.  Thus, passive advection of the magnetic field is sufficient for our purposes here.

\newtext{
The Stage 2 simulations begin from the same spherical WD that the Stage 1 simulations begin with, and makes use of
our \ac{NLTE} code for HYDrodynamcal RAdiation transport, \hydra.
Since the mechanism of \ac{DDT} is not established, it is treated as a free parameter.
The DDT is initiated by mixing of 0.01 $M_\odot$ at the burning front. 
In this simulation, the DDT is triggered after burning of about $0.27 M_\odot$ corresponding to a transition density of $ 2.5 \times 10^7 \rm{g/cm^3}$.
Unlike 
some previous studies \citep{hoeflich06,fesen07}, we will not consider
off-center delayed-detonation transition
\citep{h90,h95,hoeflich2003hydra,h03b, tiara15,Telesco2015,Hoeflich2017}.
This technique produces light curves and spectra that are consistent with
normal-bright \ac{SNeIa} \citep{Hoeflich2017}.  The addition in this current
work is the
improvement of the positron transport.}

\newtext{We should note that an attempt was made to directly map the results of  Stage
1 to the 3d hydrodynamics models in \hydra.
The structure is close to hydrostatic, and  $P_{gas}$ is many order of magnitude larger than $P_{mag}$ (see Fig. 1).  
However, omitting even a small magnetic pressure caused numerical instability
on the grid-scale, crashing the simulations.
Thus, we use this
{more simplified} approach.  }

\subsubsection{Stage 2 method: \hydra}
\label{hydra}

\newtext{
 For the Stage 2 simulations we use our
\ac{NLTE} code for HYDrodynamcal RAdiation transport, \hydra.  Here we briefly
outline the modules included.  For more details, see Appendix B.
}

\newtext{
The simulations begin with a spherical hydrostatic C/O WD, identical to that of
Stage 1
\citep{Hoeflich2017}. 
The simulations utilize a nuclear network of 218 isotopes during the early
phases of the explosion; detailed, time-dependent non-LTE models for atomic
level populations;   and {$\gamma-$ray} and positron transport and
radiation-hydrodynamics to calculate low-energy LCs and spectra
\citep{h95,hoeflich2003hydra,penney14}.
}

{The results use spherical hydrodynamics until 10 days after the explosion.
During the ongoing deflagration phase, the rate of burning is parameterized based on physical flame 
models and calibrated to 3D hydrodynamical models by A. Khokhlov 
\citep{2000ApJ...528..854D,khokhlov97,gamezo04}.
The  decay of $^{56}\rm{Ni}$ increases the specific
energy corresponding to a velocity of $\approx
 3{,}000\unit{km}\unit{s^{-1}}$ over the course of its half life,  6.1 days.
This ongoing energy input requires the hydrodynamics and the time-dependent
radiation-transport equations to be solved simultaneously in order to properly
treat cooling by the $P dV$ term.
For this early phase we solved the radiation transport equations in co-moving frames 
using $\approx 100$ frequency groups. The simulation uses atomic models with
$\approx 10^6 $ line transitions taken from the compilations of
\citet{kurucz93}, supplemented by forbidden line transitions of iron group
elements in the lower ionization states \citep{Telesco2015}. 
}

\newtext{
After 10 days,
the hydrodynamical evolution is assumed to be homologous expansion, $v \propto
r$. This
adequate because the kinetic energy dominates all other energy sources, and the sound
speed is many orders of magnitude smaller than the expansion velocity. 
}

For computational efficiency, we did not recalculate the
photosphere and transitional phase of expansion but skipped to the nebular phase
starting at day 100.

Magnetic fields impose the necessity of 3D treatment for the
positron-transport and, with it, for the 3D radiation transport for
both high and low-energy photons. For Stage 2 simulations discussed here,
we use the 3D Monte-Carlo modules for the  photon and
positrons. 

 Because of the assumptions above, we use angular averaged departure coefficients, i.e. the relative population number of the atomic levels relative to its ground state within each ion. This is
justified because the \ac{IR} radiation originates from optically thin layers,
and luminosities become isotropic. We assume stationary radiation transport,
i.e. we neglect the implicit time-dependence in the transport terms for photons
and positrons, and in the rate equations. For details, see Appendix B.

We restrict our discussion to day 1000 because after that,
additional radioactive isotopes (e.g. $^{57}{Co},^{55}Fe, ^{44}{Ti}$) are
expected to dominate the energy input, giving a natural limit for our study.

\section{Stage 1 evolution: morphology and burning rate during the early deflagration phase}
\label{sec.evolution}

\begin{figure*} \begin{center}
\includegraphics[width=0.19\textwidth]{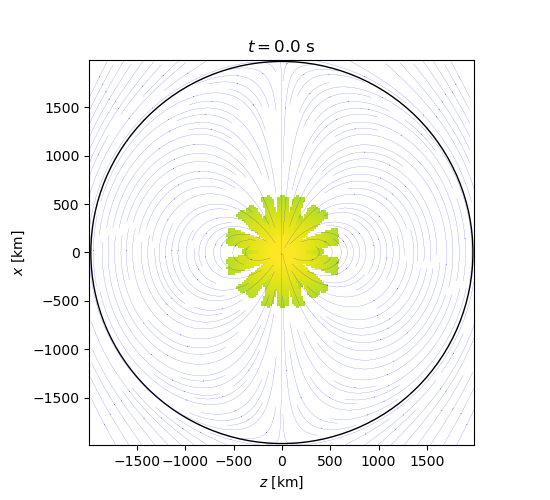}
\includegraphics[width=0.19\textwidth]{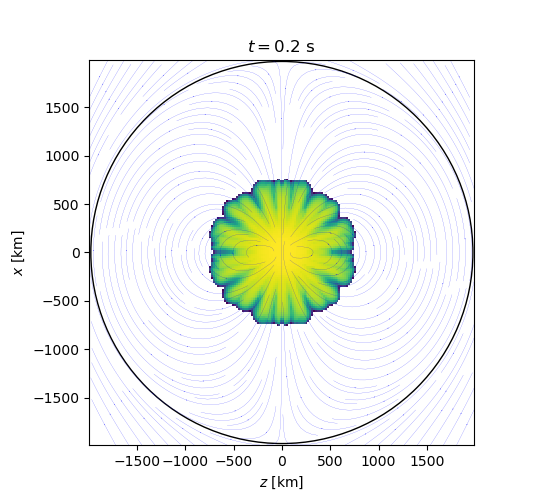}
\includegraphics[width=0.19\textwidth]{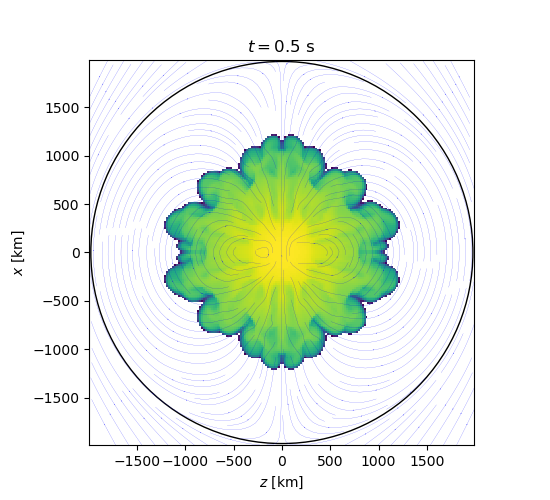}
\includegraphics[width=0.19\textwidth]{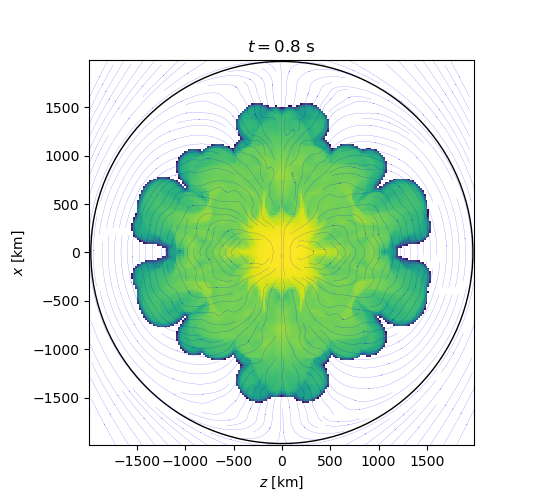}
\includegraphics[width=0.19\textwidth]{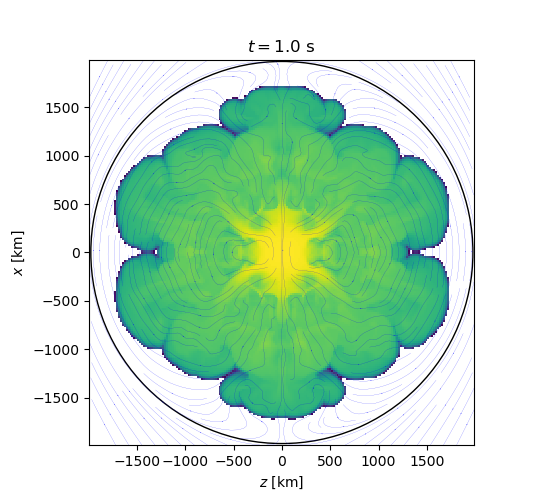}
\includegraphics[width=0.19\textwidth]{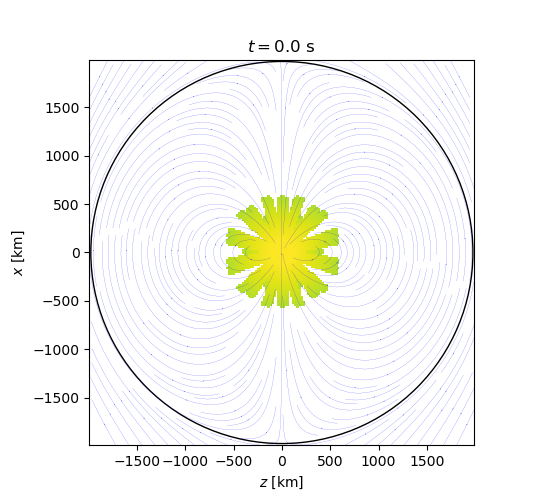}
\includegraphics[width=0.19\textwidth]{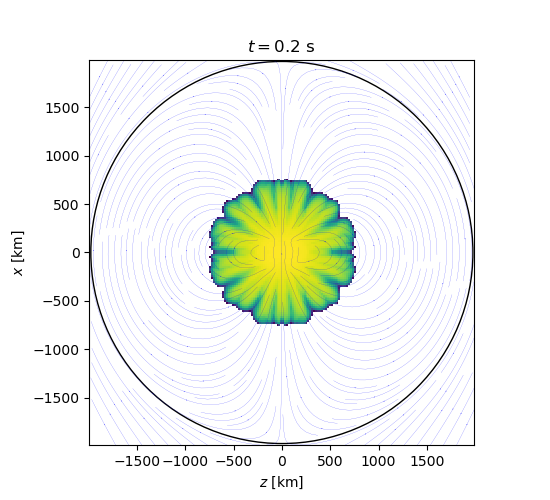}
\includegraphics[width=0.19\textwidth]{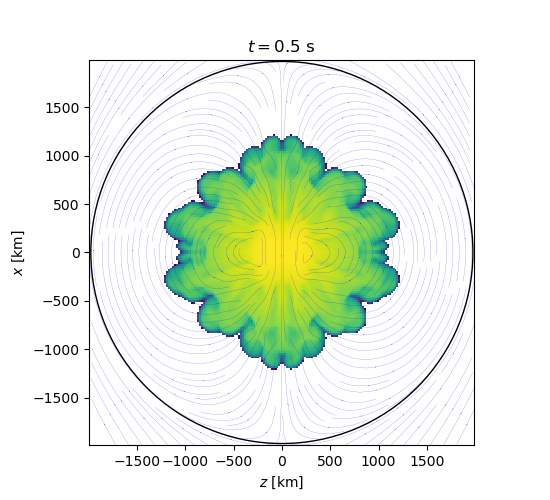}
\includegraphics[width=0.19\textwidth]{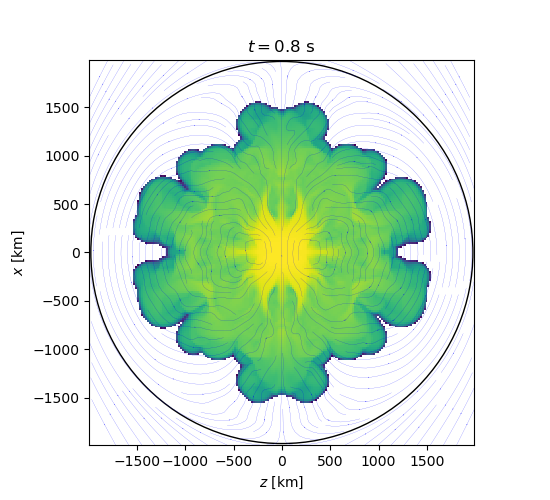}
\includegraphics[width=0.19\textwidth]{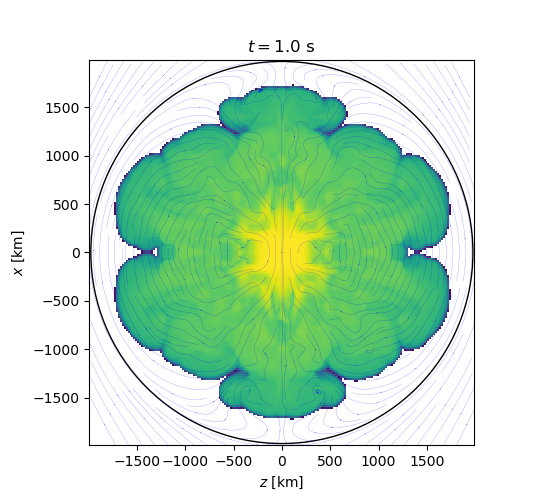}
\includegraphics[width=0.19\textwidth]{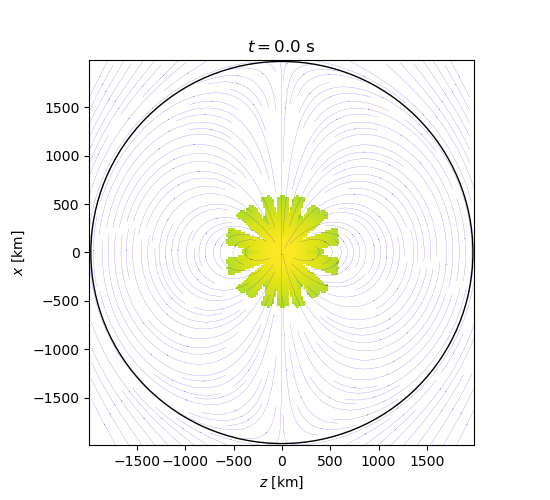}
\includegraphics[width=0.19\textwidth]{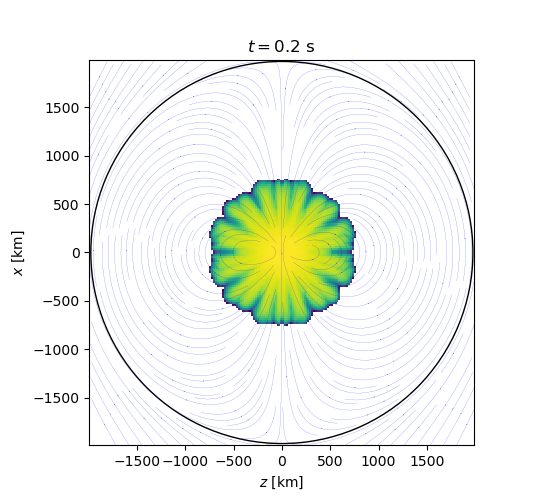}
\includegraphics[width=0.19\textwidth]{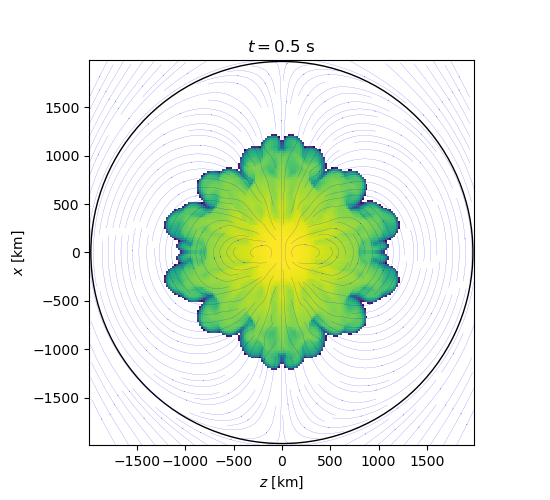}
\includegraphics[width=0.19\textwidth]{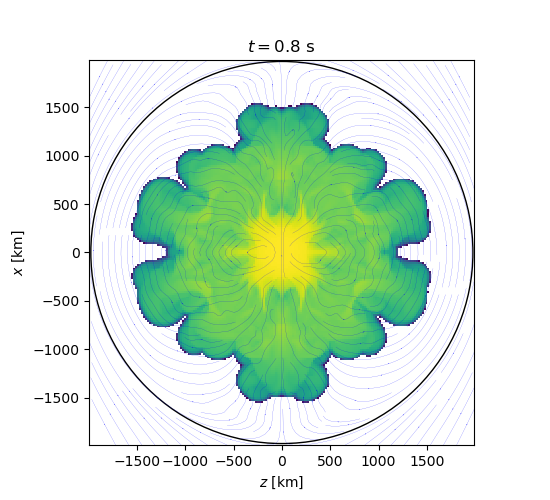}
\includegraphics[width=0.19\textwidth]{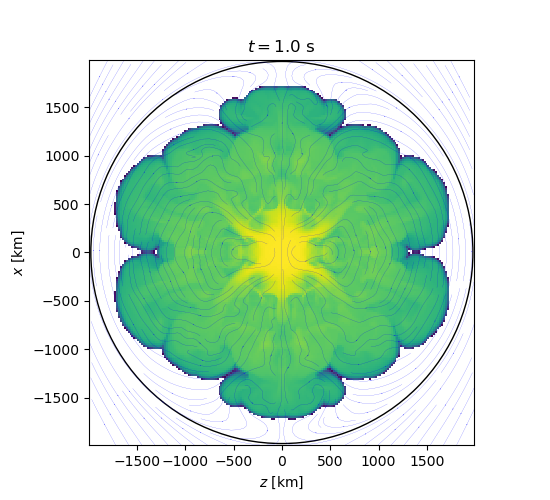}
\includegraphics[width=0.19\textwidth]{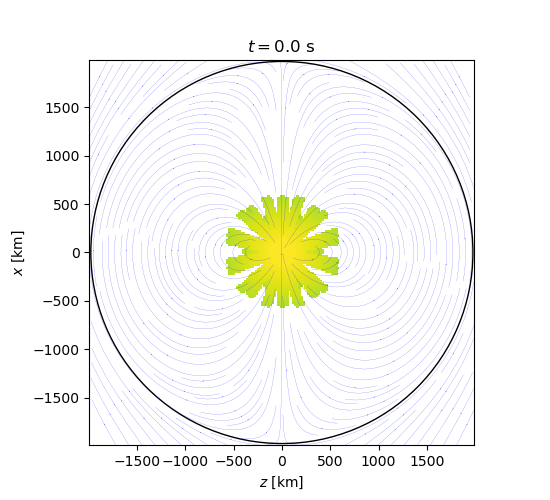}
\includegraphics[width=0.19\textwidth]{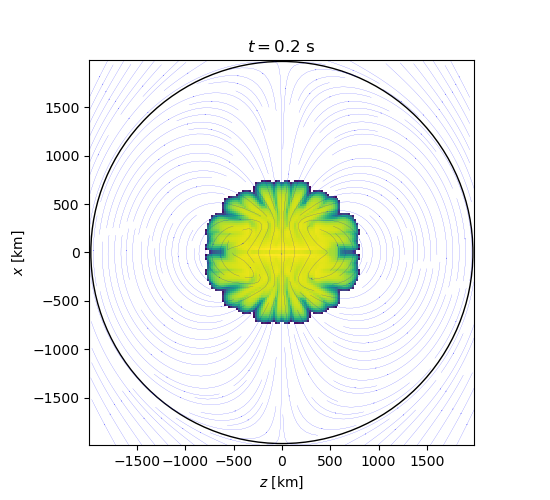}
\includegraphics[width=0.19\textwidth]{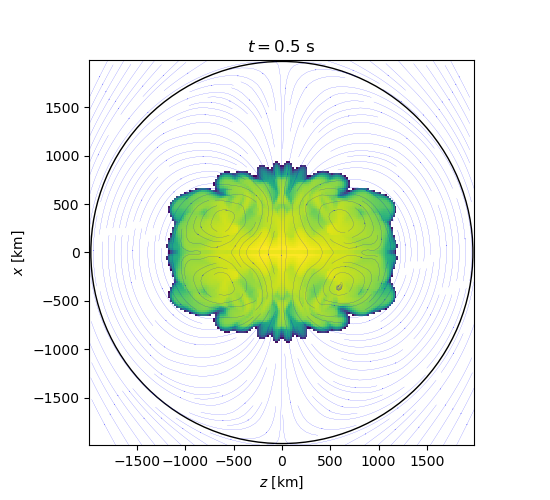}
\includegraphics[width=0.19\textwidth]{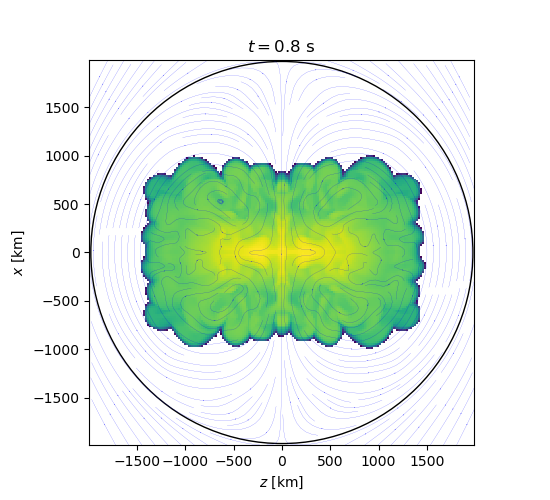}
\includegraphics[width=0.19\textwidth]{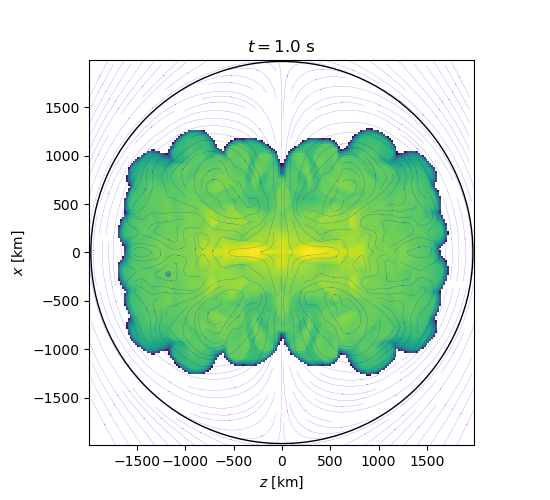}
\caption
{\textbf{Slices of Nickle Density} for all four simulations.  Magnetic
field strength increases downward, and time increases to the right.    Only the
most strongly magnetized case shows appreciable modification to the flow
morphology.}
\label{fig.ni_y} \end{center} \end{figure*}

\def\hw{0.49}
\begin{figure*}[h]
    \includegraphics[width=\hw\textwidth]{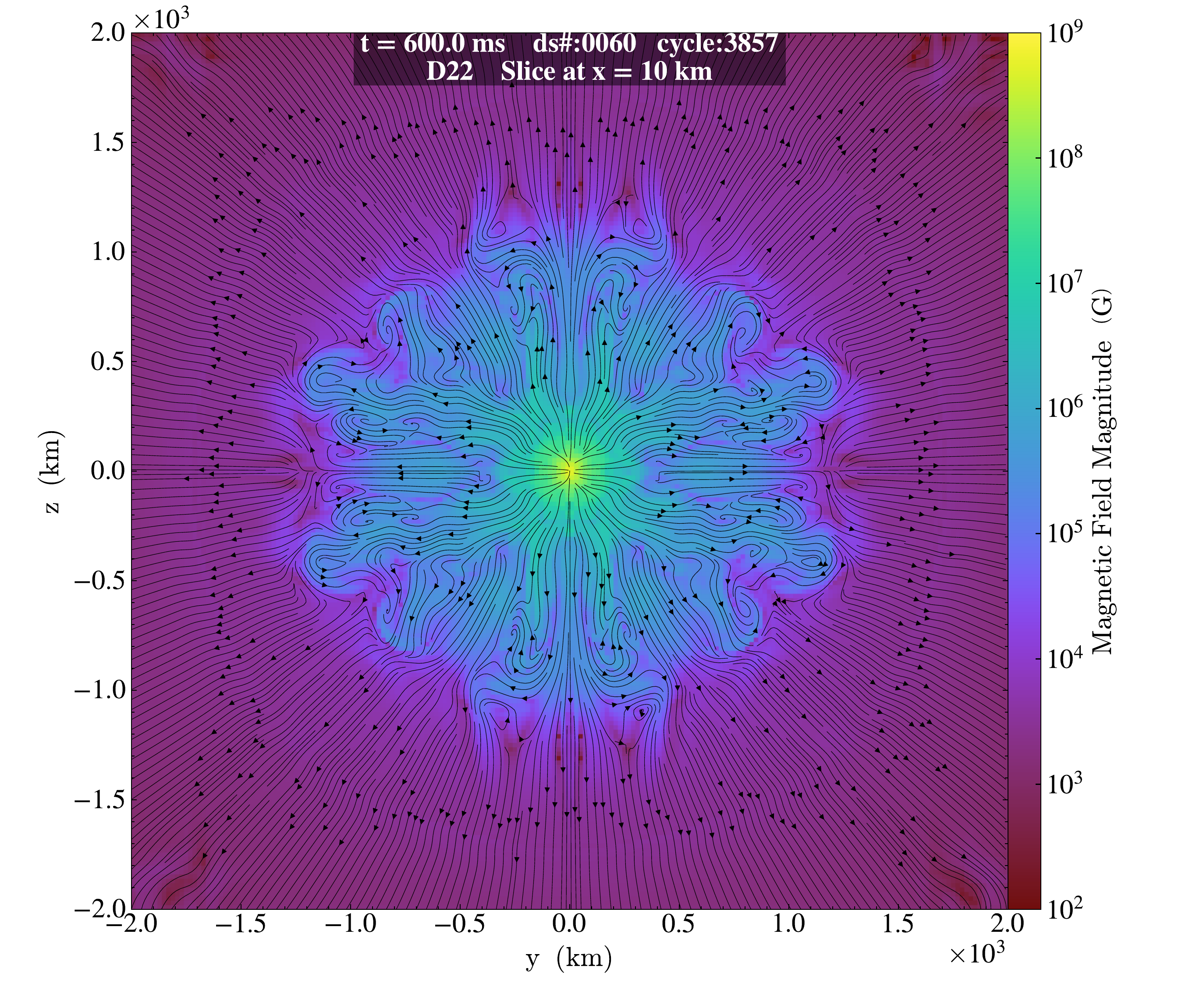}
    \includegraphics[width=\hw\textwidth]{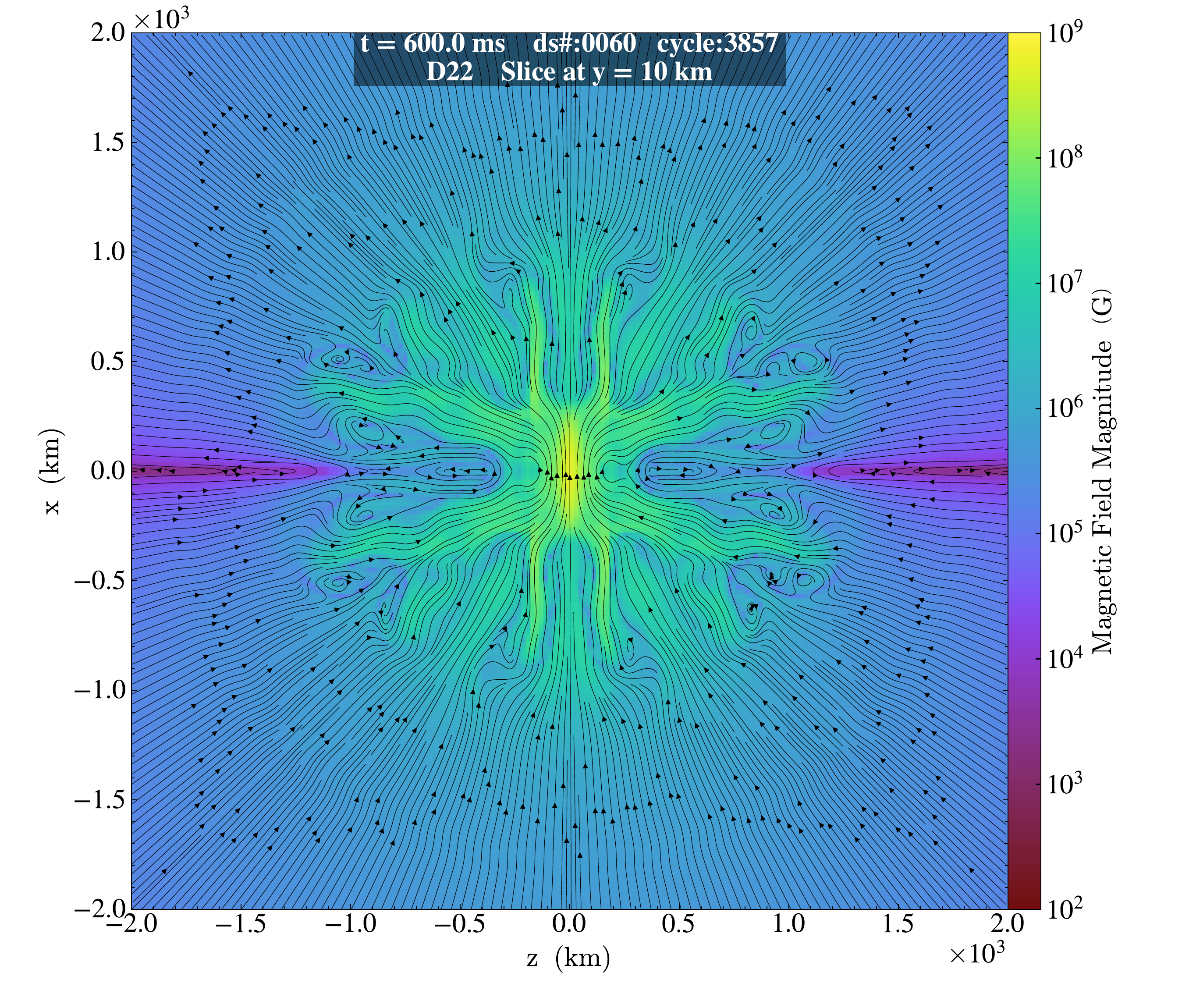}

    \includegraphics[width=\hw\textwidth]{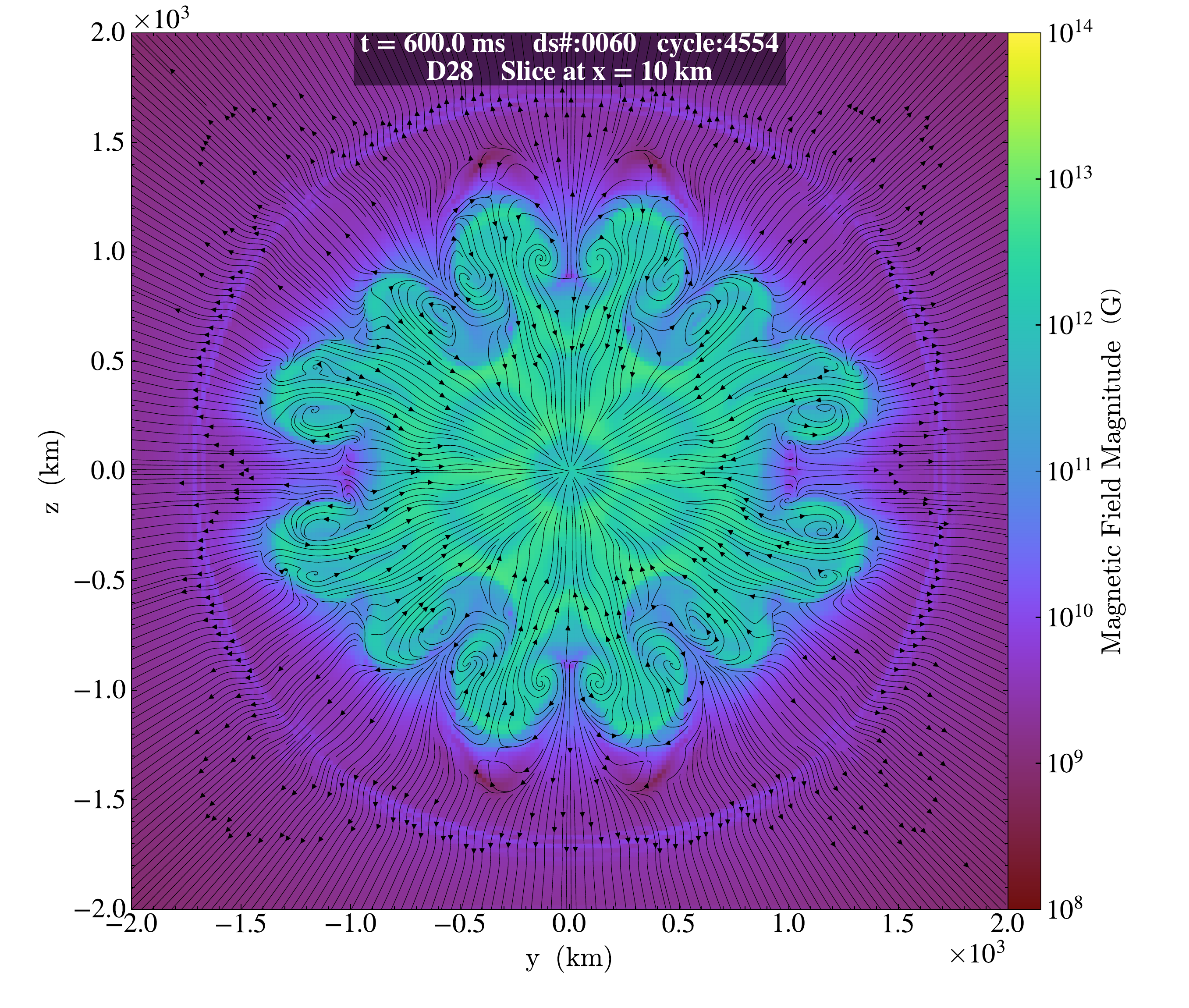}
    \includegraphics[width=\hw\textwidth]{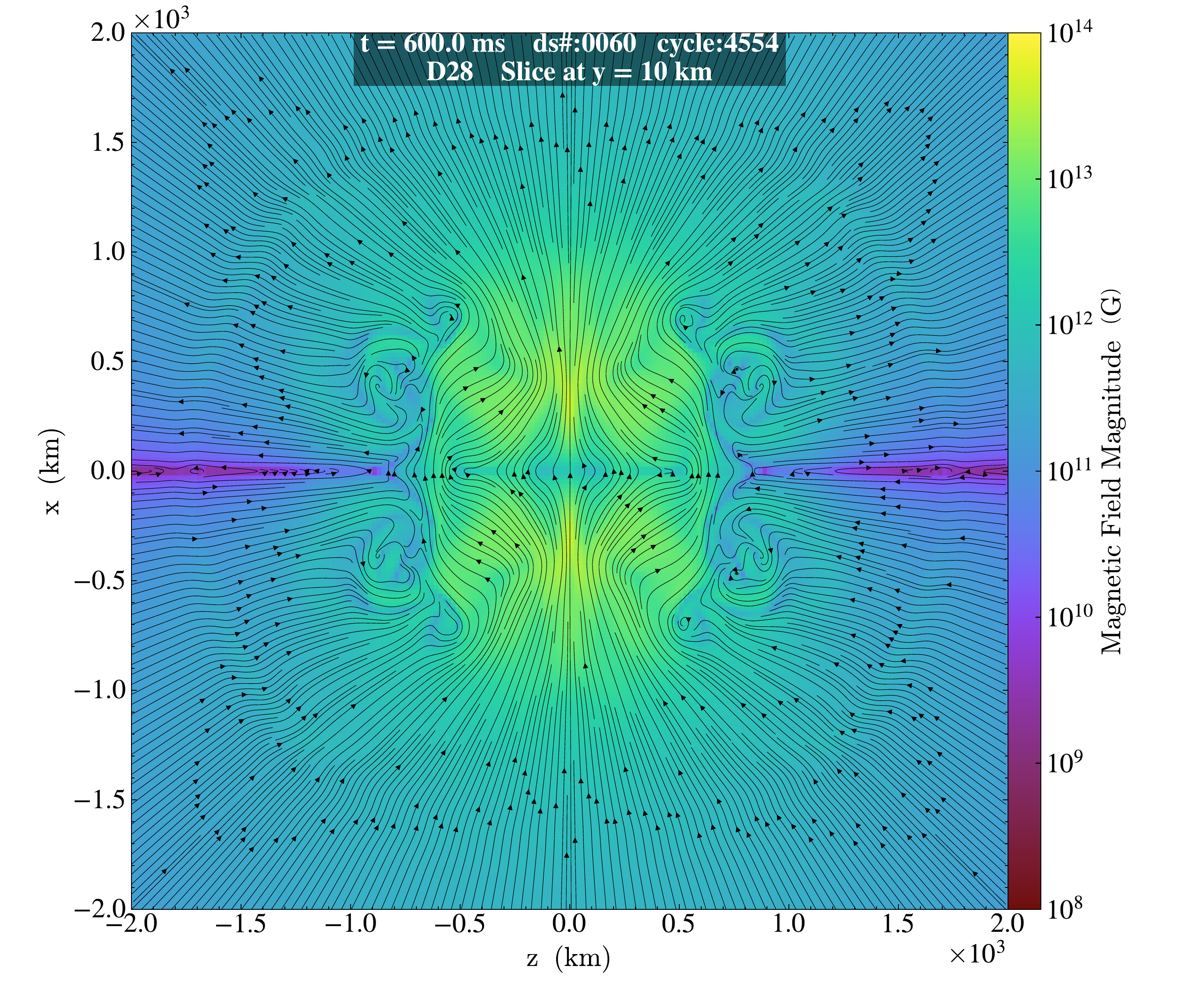}

\caption{\textbf{Magnetic Field Strength and Streamlines.}  In our weak-$B$ model, D22 (top row), the magnetic field develops eddies following the \acf{RT} instabilities. The structure of the initial global magnetic dipole is almost lost. It requires a much higher initial $B$, as in D28 (bottom row), for it to survive the \ac{RT} instabilities. Since the latter magnitudes are unrealistic, this {agrees with} the lack of observational evidence for directional dependence, suggested by models in  non-DDT scenarios.}
\label{fig.B_slices}
\end{figure*}

\begin{figure}[h] \begin{center}
\includegraphics[width=0.51\textwidth]{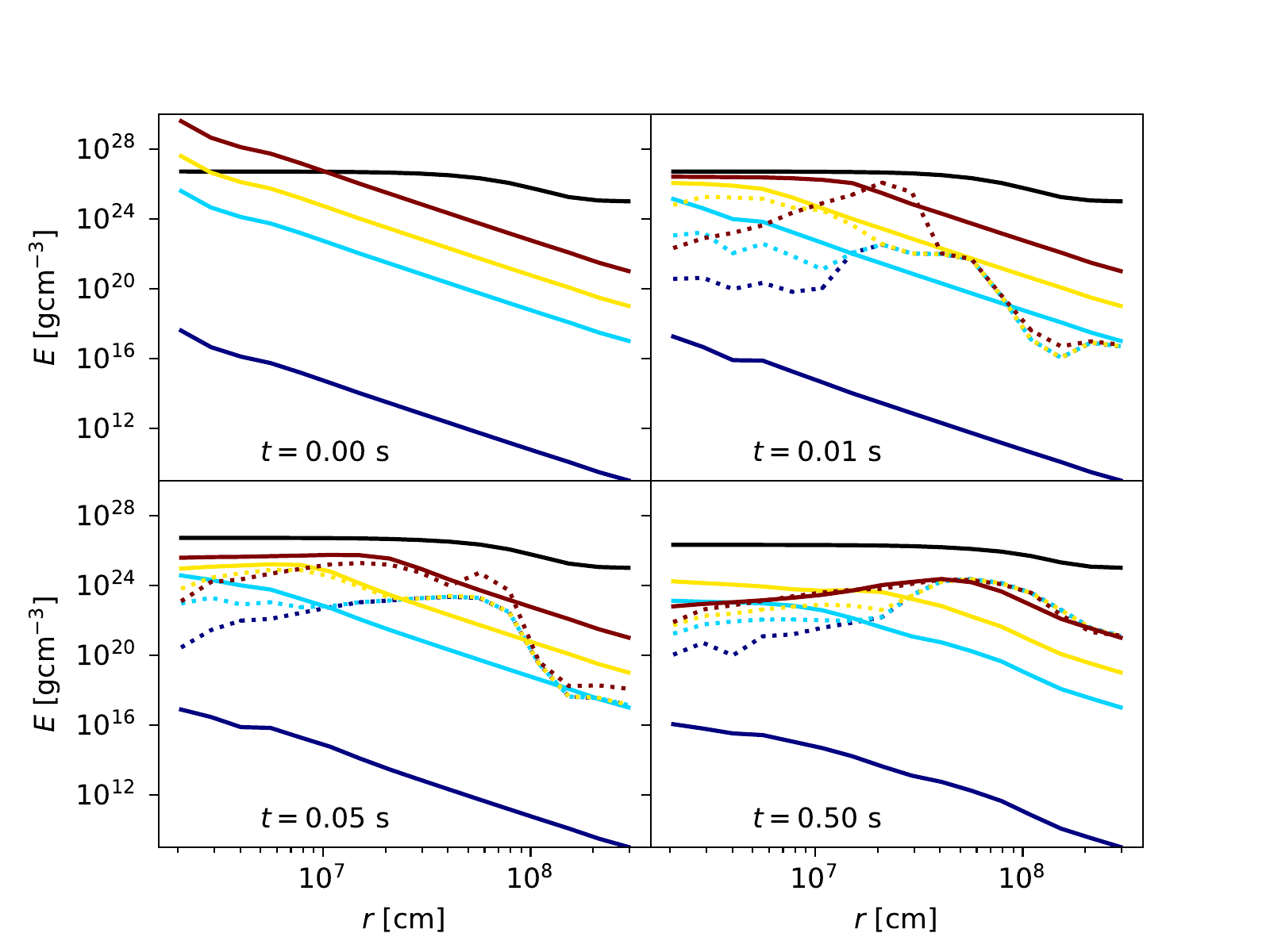}
\caption{\textbf{Energy vs. Radius} for the first $0.5\rm{s}$.  Here we show magnetic energy
density (solid lines), gas pressure (black solid lines) and kinetic energy
density (dotted lines) averaged over the sphere.  Time increases to the right.
The simulation $D22,\ D26,\ D27$ and $D28$ are color coded as red,
yellow, cyan, and blue, respectively, as in Figure 1. 
As the pressure profiles are nearly identical, we plot only one for clarity 
}
\label{fig.energy} \end{center} \end{figure}

\subsection{Burning and Rayleigh-Taylor}

The evolution of the nickel density for each of our four simulations can be seen
in Figure \ref{fig.ni_y}.  The top row of figures has the weakest magnetic
field, while the bottom has the strongest.  Time increases to the right, from
the initial condition to $t=0.6\rm{s}$.  In each simulation, the burning
proceeds outwards, retaining an imprint of the initial perturbation.  
It is notable that the magnetic field has little impact on the nickel
distribution for any but the most strongly magnetized simulations.

Gas heated by the burning subsequently develops rising fingers of magnetized gas
due to the Rayleigh-Taylor instability.  
This is seen in the magnetic field distributions in Figure
\ref{fig.B_slices}, which shows slices in the most weakly and most strongly
magnetized simulations,
$D22$ and $D28$ respectively, at $t=0.6\rm{s}$.  Due to flux freezing, the magnetic
energy that is initially strongest in the center is dragged upwards with the heated
``low'' density gas.  
The instability, and the correlation between the magnetic
field and the low density gas, is most evident in the strongly magnetized run,
see the bottom left panel of Figure \ref{fig.B_slices}.
This is counter to our initial expectation, which was that the strongest
magnetic field would suppress the Rayleigh-Taylor instability
\citep{chandra61book,Hristov2018}. Here Rayleigh-Taylor seems to be  more pronounced
in the most strongly magnetized run.

\def\ta{\ensuremath{t=0\rm{s}}}
\def\tb{\ensuremath{t=0.01\rm{s}}}
This extra \ac{RT} can be explained in part by the extra energy released during the
settling of the initial conditions.  Figure \ref{fig.energy} shows the 
energy densities for each of the four simulations (with red, yellow, cyan, and blue in
order of decreasing magnetic field strength). {Black solid lines show the gas pressure, color} solid lines show magnetic energy,
color {dotted} lines show kinetic energy.  The pressure profile for each can
be seen as the black dashed line.  The pressure profile evolves very little
during this evolution, and is nearly the same for each of the simulations.  At
\ta, the magnetic energy is below the gas pressure for all radii for the two
weakly magnetized fields.  The two more strongly magnetized runs have a small
excess of magnetic energy. The excess magnetic energy is releases in the first
0.01s, and in turn drives 
a somewhat larger flow than that seen in the other two simulations.  It can also
be seen that the strongly magnetized run, $D28$, the magnetic and kinetic
energies are roughly balanced over all radii by $t=0.05\rm{s}$.  For the other
simulations, kinetic energy due to the burning outweighs magnetic energy at
outer radii.  
It is interesting to note that even though the magnetic energy is large in 
$D28$, there is still clearly defined Rayleigh-Taylor.  This shows that
equipartition of magnetic and kinetic energies is not a sufficient criterion for
suppression of hydrodynamical instabilities.

It can be seen in the bottom row of Figure \ref{fig.ni_y} that the burning
Nickel in the most magnetized run is flattened along the pole-ward direction.
This is due mainly to the anisotropic magnetic pressure, which is larger along
the poles.  Only in D28 is this pressure large enough to alter the burning.

\begin{figure}
\centering
\includegraphics[width=0.49 \textwidth]
{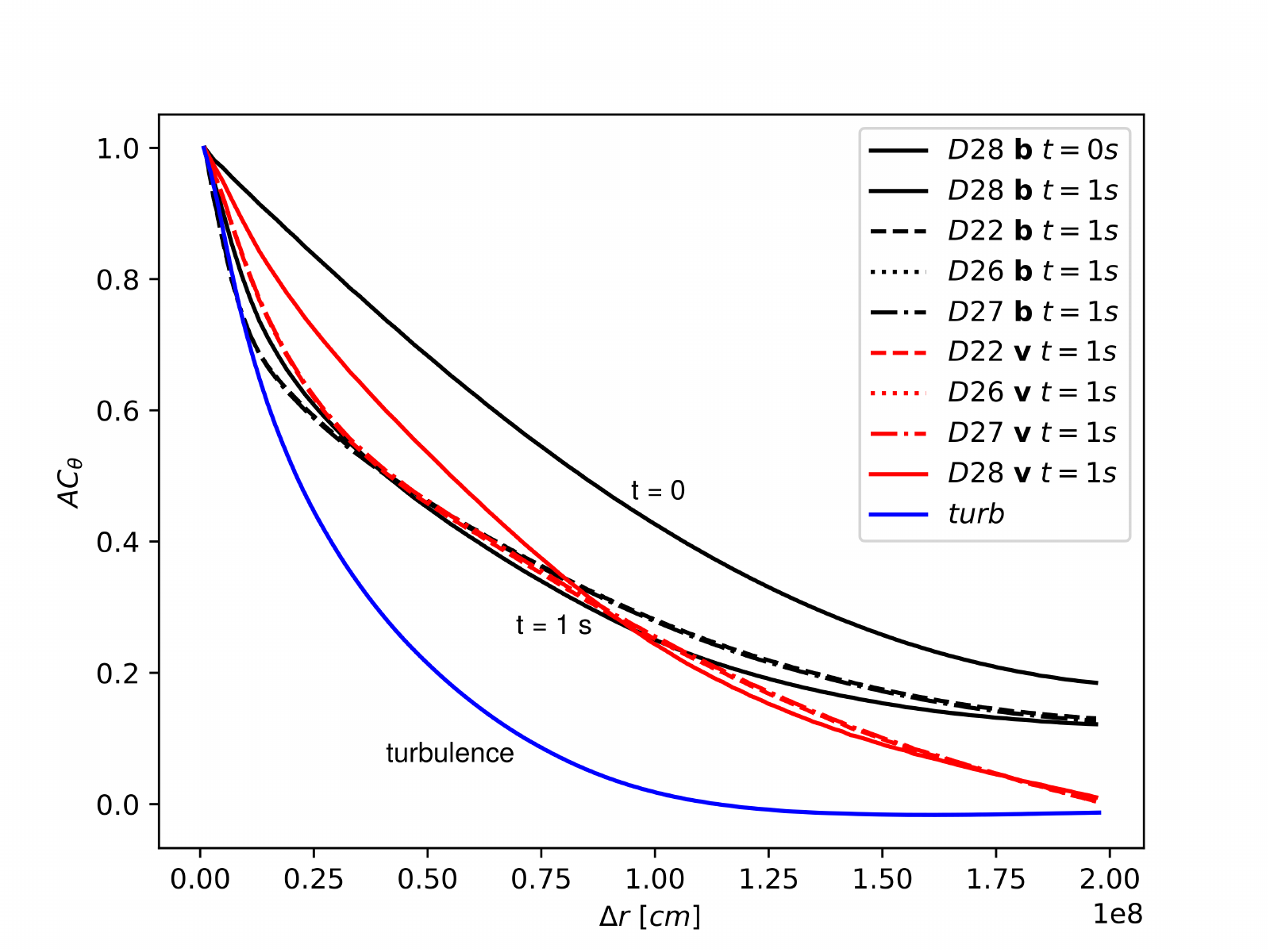}
\caption{ \textbf{The Auto Correlation Function} for the magnetic field alignment. Black
curves show the correlation in the magnetic field direction, $AC_{\theta,B}$.
The velocity direction auto-correlation, $AC_{\theta,v}$, is shown in red.
As expected, correlations in the magnetic field are determined by correlations
in the velocity for all but the most strongly magnetized simulations.  In the
most strongly magnetized case, the velocity is more correlated, and the field is
less correlated than the velocity.  The blue line shows $AC_{\theta,B}$ for a simulation of driven
turbulence, to indicate an extreme case of disorder.  The correlation in the field itself is intermediate between an
ordered dipole and a fully turbulent field.}
\label{fig.ac}\end{figure}
\begin{figure}[h] \begin{center}
\includegraphics[width=\hw\textwidth]{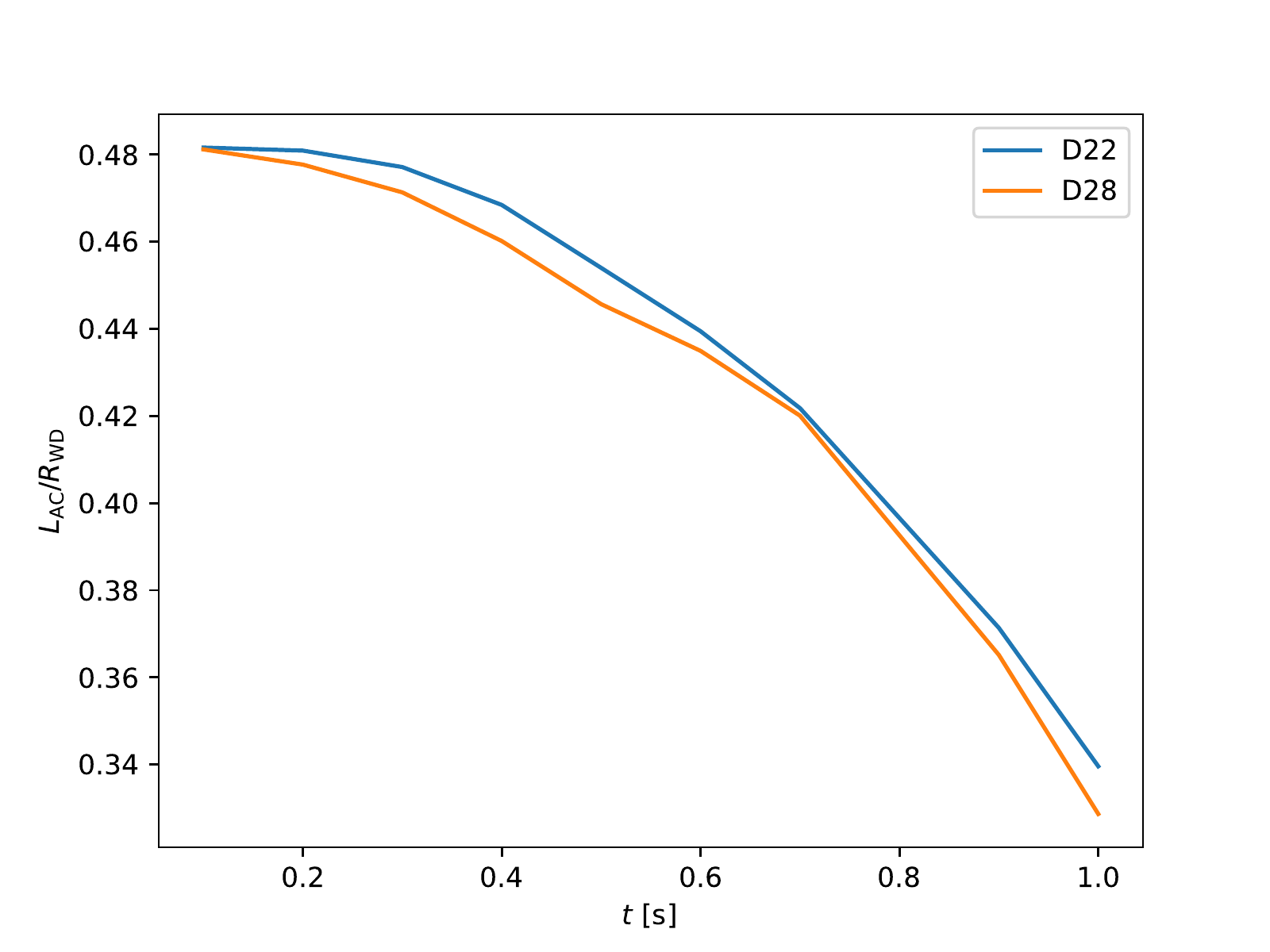}
\caption{\textbf{The auto-correlation length}, defined by eq. (\ref{eq.L_ac}), of the magnetic field as a function of
time, in units of the initial white dwarf radius, $R_{\rm{WD}}$.  Initially, the
magnetic field is correlated over {half the} star, but decreases to a fraction of
that.  Improved resolution would likely cause this to decrease further.}
\label{fig.ac_time} \end{center} \end{figure} 

\def\rvec{\ensuremath{\mathbf{r}}}
\def\vvec{\ensuremath{\mathbf{v}}}
\def\vvech{\ensuremath{\hat{\vvec}}}
\def\bvec{\ensuremath{\mathbf{b}}}
\def\Bvec{\ensuremath{\mathbf{B}}}
\def\bvech{\ensuremath{\hat{\bvec}}}
\def\xvec{\ensuremath{\mathbf{x}}}
\def\acb{\ensuremath{AC_{\theta,B}}}
\def\acv{\ensuremath{AC_{\theta,v}}}

\subsection{Tangling}

The structure of the magnetic field, while initially a dipole, is quickly tangled
by the kinetic motions of the gas.  This can be seen clearly in Figures
\ref{fig.B_slices} and \ref{fig:streamlines},
which shows the magnetic field lines 
 for the two extreme cases.  The magnetic field structures clearly follow
the density structures, becoming substantially more tangled. This is due to the
action of flux freezing, whereby magnetic flux is pinned to the moving fluid.  

In order to quantify the degree of tangling in the magnetic and velocity fields, we use the auto-correlation
functions
$AC_{\theta,\myhatbf{b}}(\Delta x)$ and $AC_{\theta,\myhatbf{v}}(\Delta x)$ 
of the respective direction vectors,
$\myhatbf{b} = \mathbf{B} / B$ and $\myhatbf{v} = \mathbf{v} / v$:
    \begin{align}
    AC_{\theta,B}(\Delta x) = 
    \frac{1}{4 \pi (\Delta x)^2 V}
    \int_{S(\mathbf{x}, \Delta x)} ds(\Delta \mathbf{x}) 
    \int d^3 \mathbf{x} \ 
    \myhatbf{b}(\mathbf{x}+\Delta \mathbf{x}) \cdot \myhatbf{b}(\mathbf{x}),
\\
    AC_{\theta,v}(\Delta x) = 
    \frac{1}{4 \pi (\Delta x)^2 V}
    \int_{S(\mathbf{x}, \Delta x)} ds(\Delta \mathbf{x}) 
    \int d^3 \mathbf{x} \ 
    \myhatbf{v}(\mathbf{x}+\Delta \mathbf{x}) \cdot \myhatbf{v}(\mathbf{x}).
\end{align}
\newtext{These functions measure the degree to which the fields $\myhatbf{v}$ and
$\myhatbf{b}$ are correlated with themselves at a separation of $\Delta
\mathbf{x}$.  
The interior integration is an average over the domain $V$. 
The outer integration is over the direction of the shift
$\Delta x$, so the functions depend only on the magnitude of the shift and not its direction. The above function are normalized  so that $AC(0) = 1$.  
At zero separation, anything is perfectly correlated with itself. A field with
more spatial disorder, and hence increased tangling of the fields, will show a
lower correlation with itself at a certain distance.  
Figure \ref{fig.ac} shows the auto-correlation function for our dipole (labeled
$t=0$) which shows a nearly linear decrease in the correlation with distance.
On the same graph, a ``maximally disordered'' configuration of driven high
Reynolds number MHD turbulence (blue line) is shown for comparison.  The more tangled turbulent state
shows lower auto-correlation, despite being statistically spatially homogeneous,
due to the chaotic nature of turbulence.  }

\newtext{Figure \ref{fig.ac} also shows the $B$ and $v$ auto-correlations at $t = 1.0 \,\rm{s}$
for all runs.  
All four simulations begin with the same dipole field.
As the simulation proceeds, the burning drives a number of instabilities, among
them Rayleigh-Taylor and Kelvin-Helmholtz, which cause fluctuation in the field
direction due to flux freezing.  This can be seen as
a decrease in the auto-correlation.  Three of the simulations have identical
auto-correlation functions.  These are shows as dashed, dotted, and dot-dashed
lines in black, but are indistinguishable from each other.  Only the most
strongly magnetized run shows any deviation.  That simulation (solid black and
red lines) shows small differences in the magnetic correlation, with the
magnetic field resisting some small amount of tangling.  It additionally shows a
substantial difference in the velocity correlation, with larger scale flows
being more active.  }

We can further quantify the tangling by defining the auto-correlation length,
\begin{align}
\label{eq.L_ac}
    L_{ac} = \int_0^{\infty}  \mathbf{d}\Delta x AC_{\theta,B}(\Delta x) /AC_{\theta,B}(0),
\end{align}
which describes the length at which there is appreciable correlation.
This quantity can be seen relative to the initial white dwarf radius in Figure
\ref{fig.ac_time}.   This plot shows that the correlation length monotonically
decreases with time, as the field becomes more tangled, and this length is not
particularly sensitive to our initial magnetic field.

\newtext{
The conclusion from this discussion is that the topology of the magnetic field 
is determined primarily by the burning front, and not greatly influenced by the
initial field.  The topology of the magnetic field, as well as its strength, influences
the rate at which positrons diffuse away from their place of origin.  
Larger correlation lengths allow positrons to travel further, while a more
tangled field traps the positrons.  As our initial conditions are heavily
idealized, we cannot predict what values $L_{ac}$ will attain for a real SNIa,
but we have established that a burning supernova cannot support a magnetic field
topology as simple as a dipole.}

Additionally, we are further justified in neglecting the back-reaction of the
magnetic field on the flame during the Stage 2 simulations.  Since the strength
of the field does not impact its topology, it also does not impact the burning
front in an appreciable way, and we can neglect it for our purposes in Stage 2.   While there is a small effect of the field in the strongly magnetized run, it is unlikely to change our conclusions in a qualitative manner.

\begin{figure}[h] \begin{center}
\includegraphics[width=0.5\textwidth]{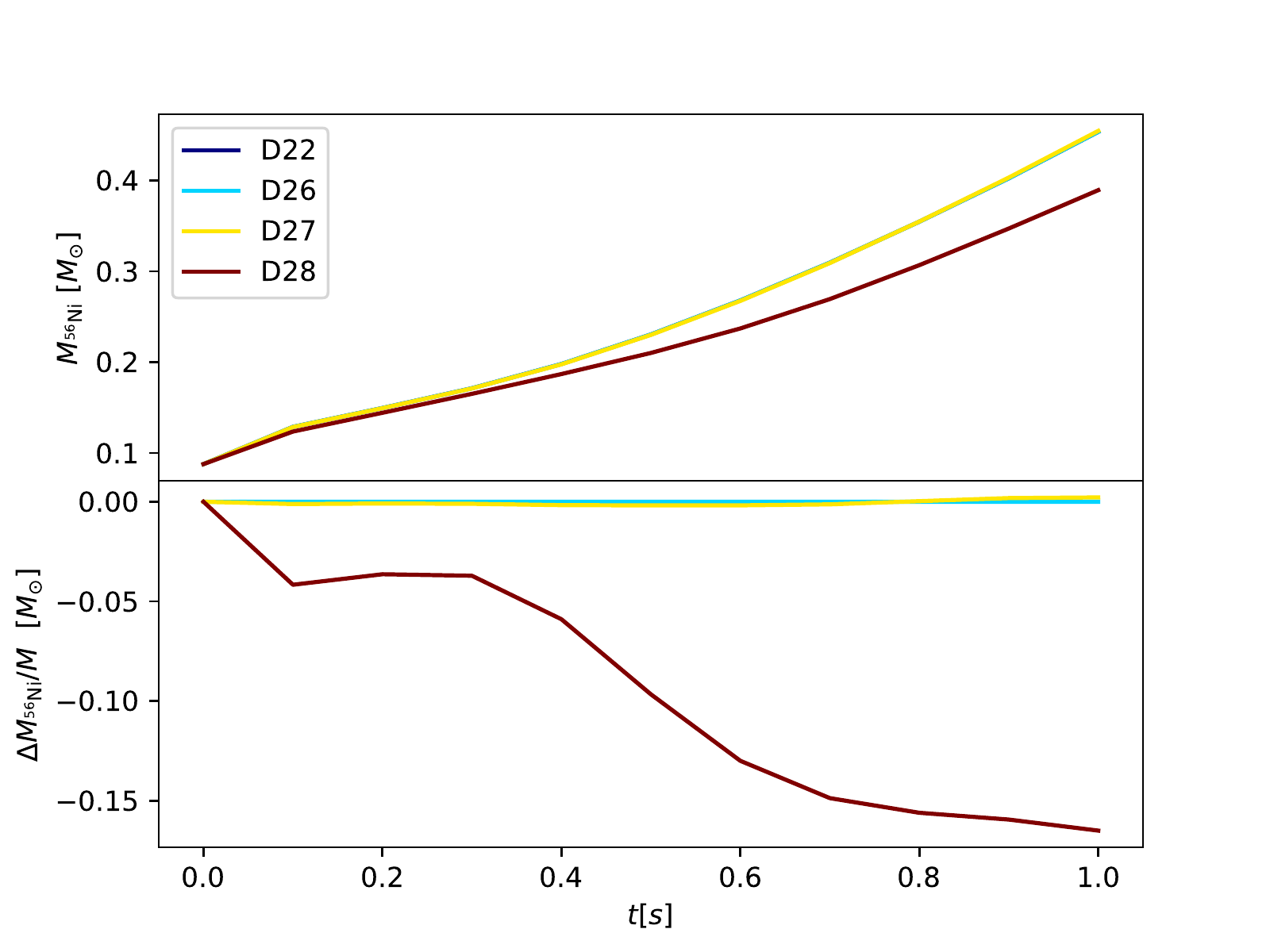}
\caption{\textbf{Nickel mass vs. time} for all four simulations \emph{(top)} and
mass difference relative to the D22.  The three weakly 
magnetized simulations are nearly indistinguishable, but the strongly magnetized
run shows a sustained reduction of the burning rate for the duration of the
simulation.}
\label{fig.nickle_time} \end{center} \end{figure}

\subsection{Burning}
We show the difference in burned mass in Figure \ref{fig.nickle_time}.  From
this, we see the strongly magnetized case, $D28$, burns about 15\%\ less nickel
than the other three runs.
For the initial drop before 0.1s, the difference in burning rate is due to the 
initial excess of magnetic energy that
decreases the central density somewhat.  
Beyond 0.1s, the difference in burning rates is largely
due to differences in the morphology of the interface.  This can be seen by
comparing the difference in burning rate (the slope of the bottom panel of
Figure \ref{fig.nickle_time}) and the difference in $L_{\rm{AC}}$ in Figure
\ref{fig.ac_time}.

\begin{figure} \begin{center}
\includegraphics[width=\hw\textwidth]{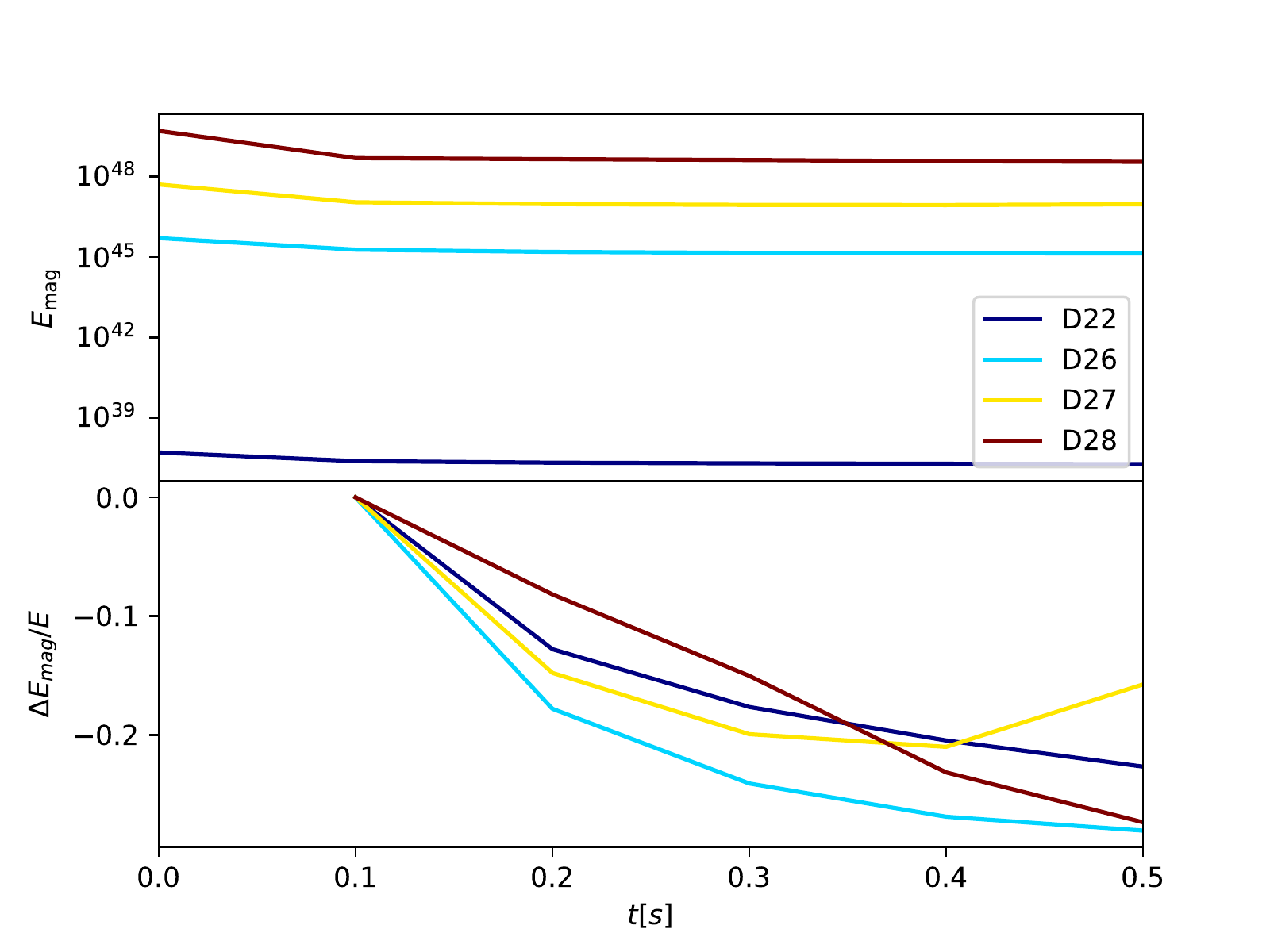}
\caption{\textbf{Magnetic energy vs. time} for each of the runs \emph{(top)}.
\emph{(Bottom)} Relative magnetic energy change after the initial conditions relax.  No evidence of field growth is seen except for a small increase in $D27$.   }
\label{fig.field_time} \end{center} \end{figure}

\subsection{Amplification}

Figure \ref{fig.field_time} shows the magnetic energy vs. time for each
simulation.  There is an initial drop as the field relaxes from our
out-of-equilibrium initial conditions, and the field energy is converted in part
to kinetic energy (see Figure \ref{fig.energy}).
  Since the burning is nearly identical, the
difference in kinetic energy between the runs comes from the release of magnetic energy.  To
examine the subsequent evolution, we plot the  change in total magnetic
energy relative to $t=0.1$s, right after the initial conditions relax.  
The
total magnetic energy for three of the runs continues to decrease, while the 
$D27$, the second-strongest field, shows a slight increase.

\begin{figure} \begin{center}
    \includegraphics[width=0.51\textwidth]{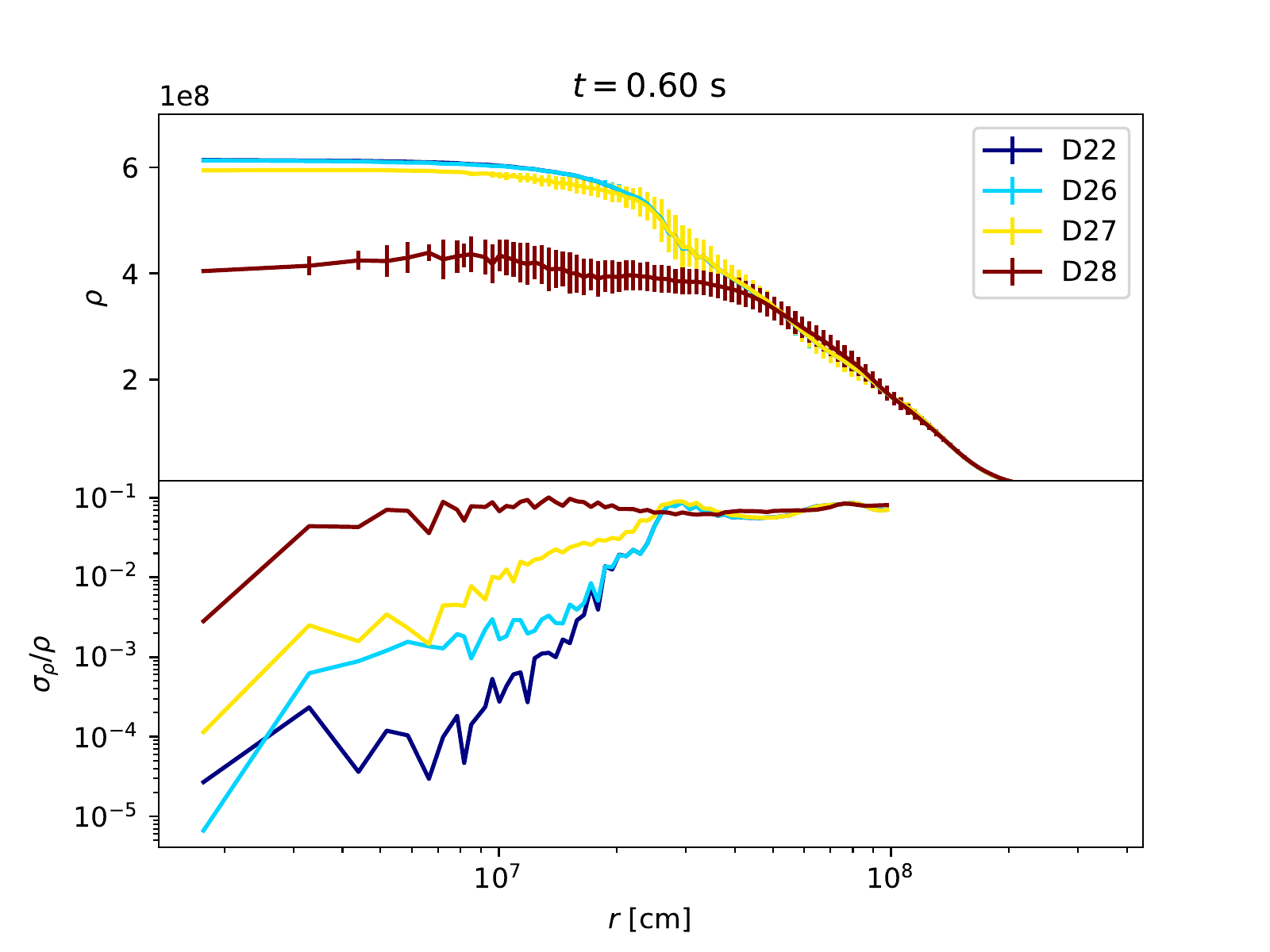}
    \caption{\textbf{Density vs. Radius} at $t=0.6\rm{s}$ for each of our
    simulations. Error bars show deviation from spherical.  \emph{(Bottom)}
    The standard deviation in the density.  Errors in the strongly magnetized
    run are on the order of a few percent.}  
\label{fig.radial_density} \end{center} \end{figure}

\subsection{Spherical Symmetry}

\newtext{Even though there is substantial magnetic field, the structures within the WD
stay relatively spherically symmetric on average because the overall system stays in hydrostatic equilibrium dominated by the gravitation. Figure
\ref{fig.radial_density} shows the angle-averaged density vs. radius {at $t = 0.6$ s.}  The top
panel shows the average density and the standard deviation from spherical,
$\sigma_\rho(r)=\sqrt{\langle(\rho(r)-\bar{\rho}(r))^2\rangle}$.  Error bars are
shown on all plots, but smaller than the line width for small radius.  The
deviation from spherical
is most pronounced in the strongly magnetized run.  The bottom panel of that
figure shows the normalized standard deviation, $\sigma_\rho/\rho$.  The peak
$\sigma_\rho$ is $8\%$ for the most magnetized case.  The few percent deviation
enjoyed by all simulations at $r>3\times 10^{7}\rm{cm}$ is due to the boundary
of the domain.  Future work will employ a slightly larger domain to avoid this
effect.  The majority of the simulations are spherical to the fraction of a
percent for the bulk of the evolution.}

\begin{figure*}[h]
    \includegraphics[width=0.49\textwidth]{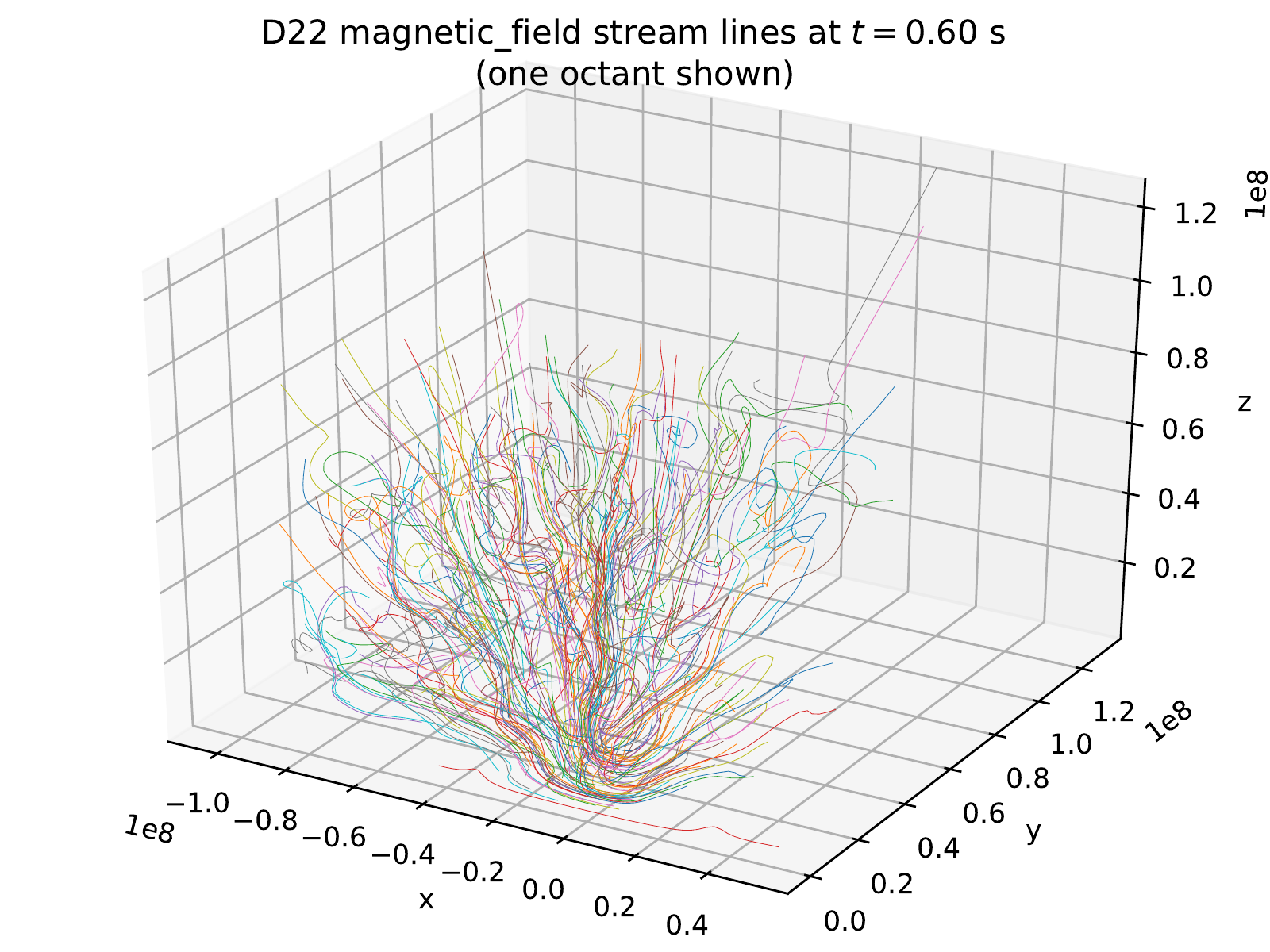}
    \includegraphics[width=0.49\textwidth]{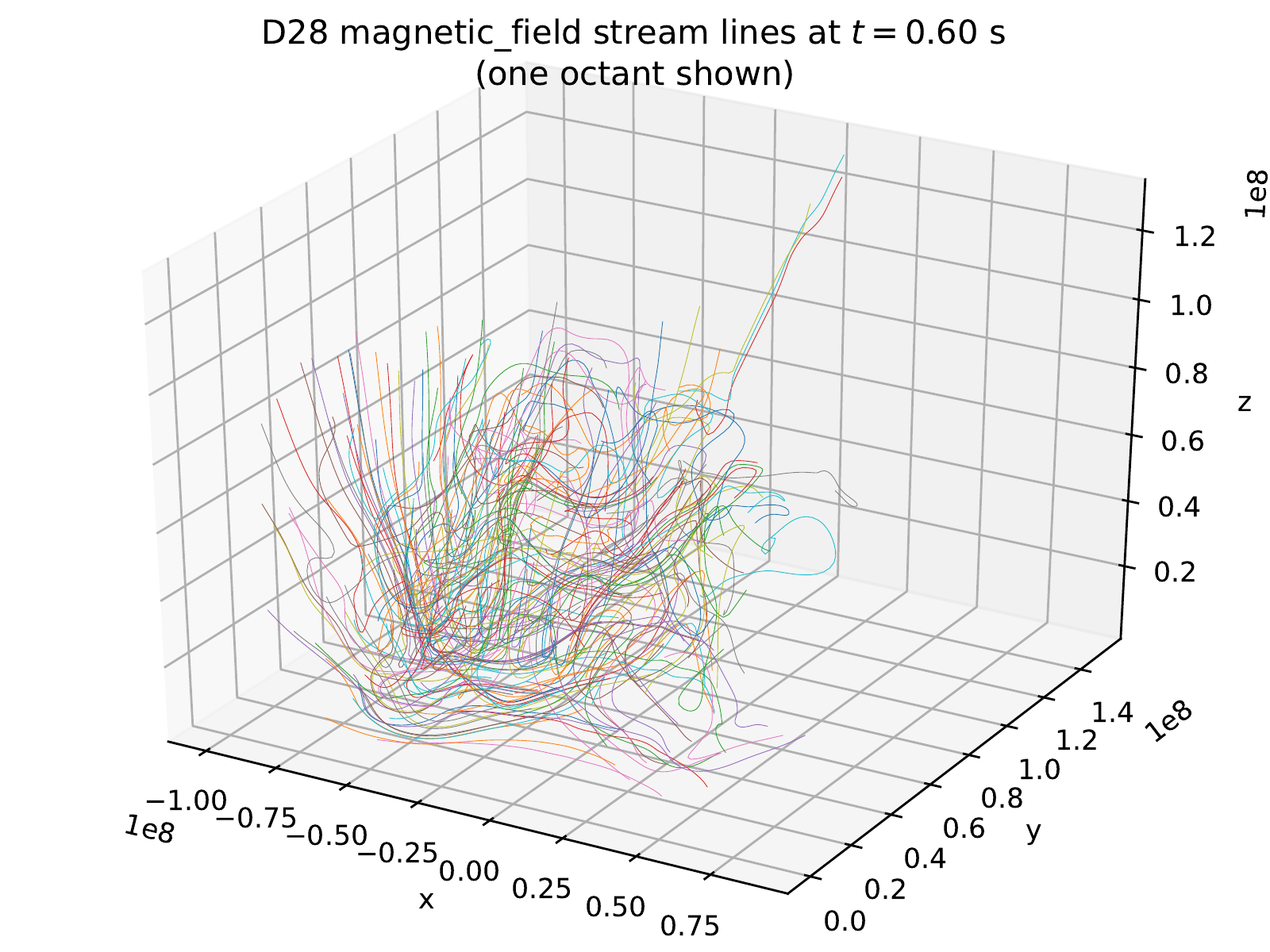}
    \includegraphics[width=0.49\textwidth]{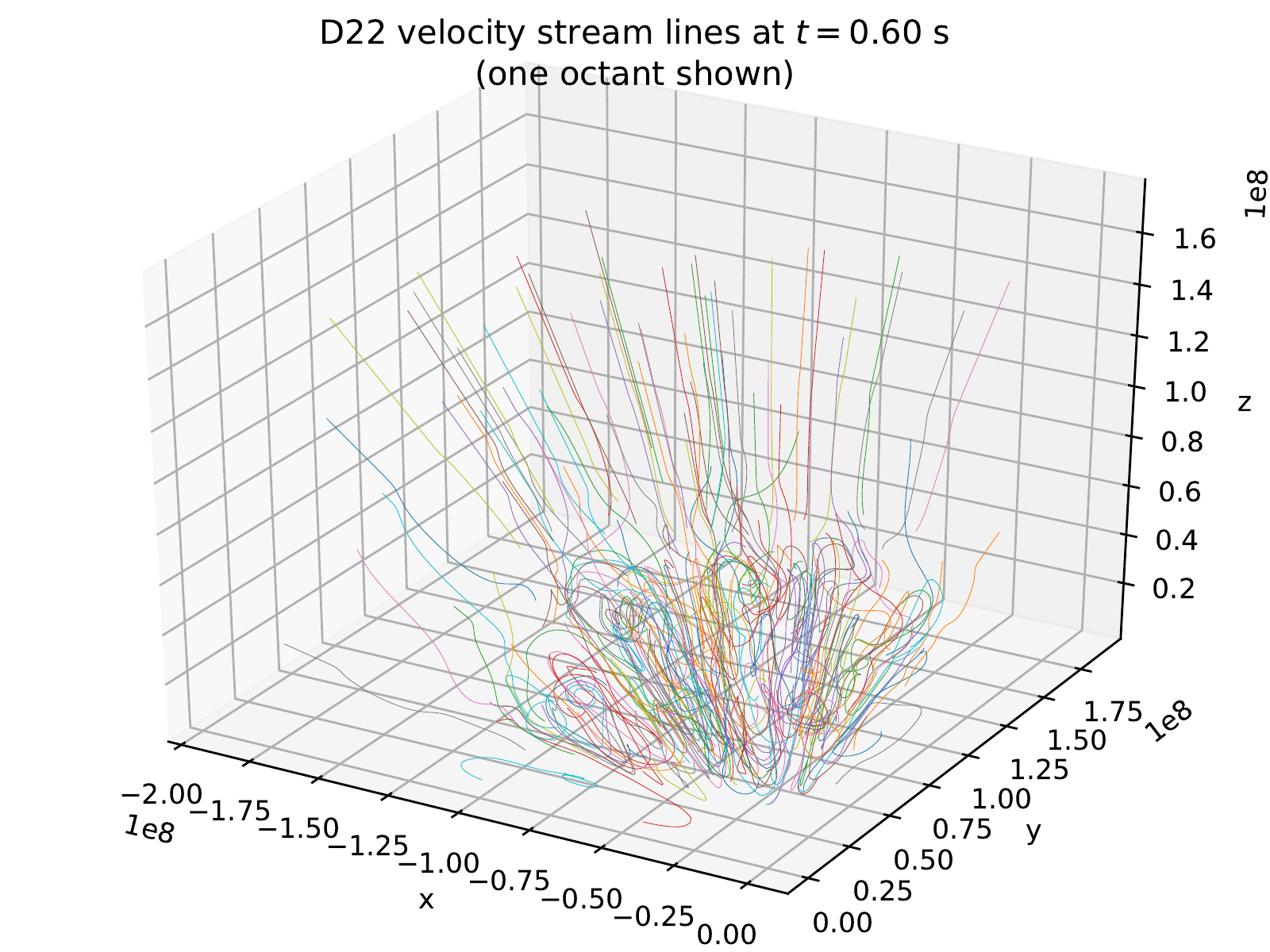}
    \includegraphics[width=0.49\textwidth]{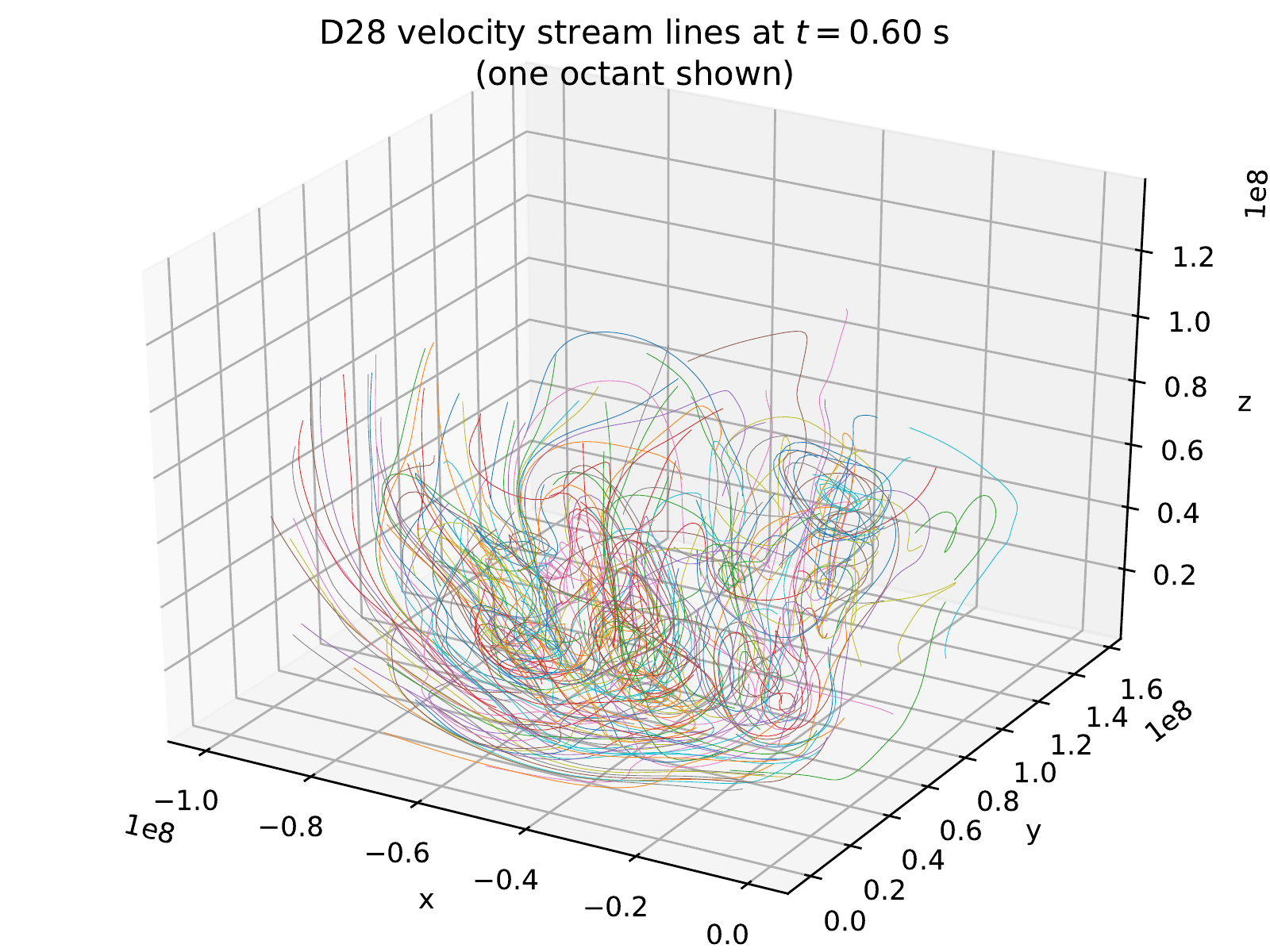}
\caption{\textbf{Magnetic field line} ``hairballs'' In our weak-$B$ model, D22
(left column), the velocity field develops \acf{RT} instabilities and the
magnetic field eddies follow. The structure of the initial global magnetic
dipole is almost lost. It requires a much higher initial $B$, as in D28 (right
column), for it to survive the \ac{RT} instabilities. Since the latter
magnitudes are unrealistic, this show why there is no observational evidence of
directional dependence, suggested by models in  non-DDT scenarios. For clarity
the figures show field lines originating in the $x<0,\  y>0,\  z>0$ octant only.
\label{fig:streamlines}
}
\end{figure*}

\begin{figure}[h]
\begin{center}
\includegraphics[width=0.45\textwidth]{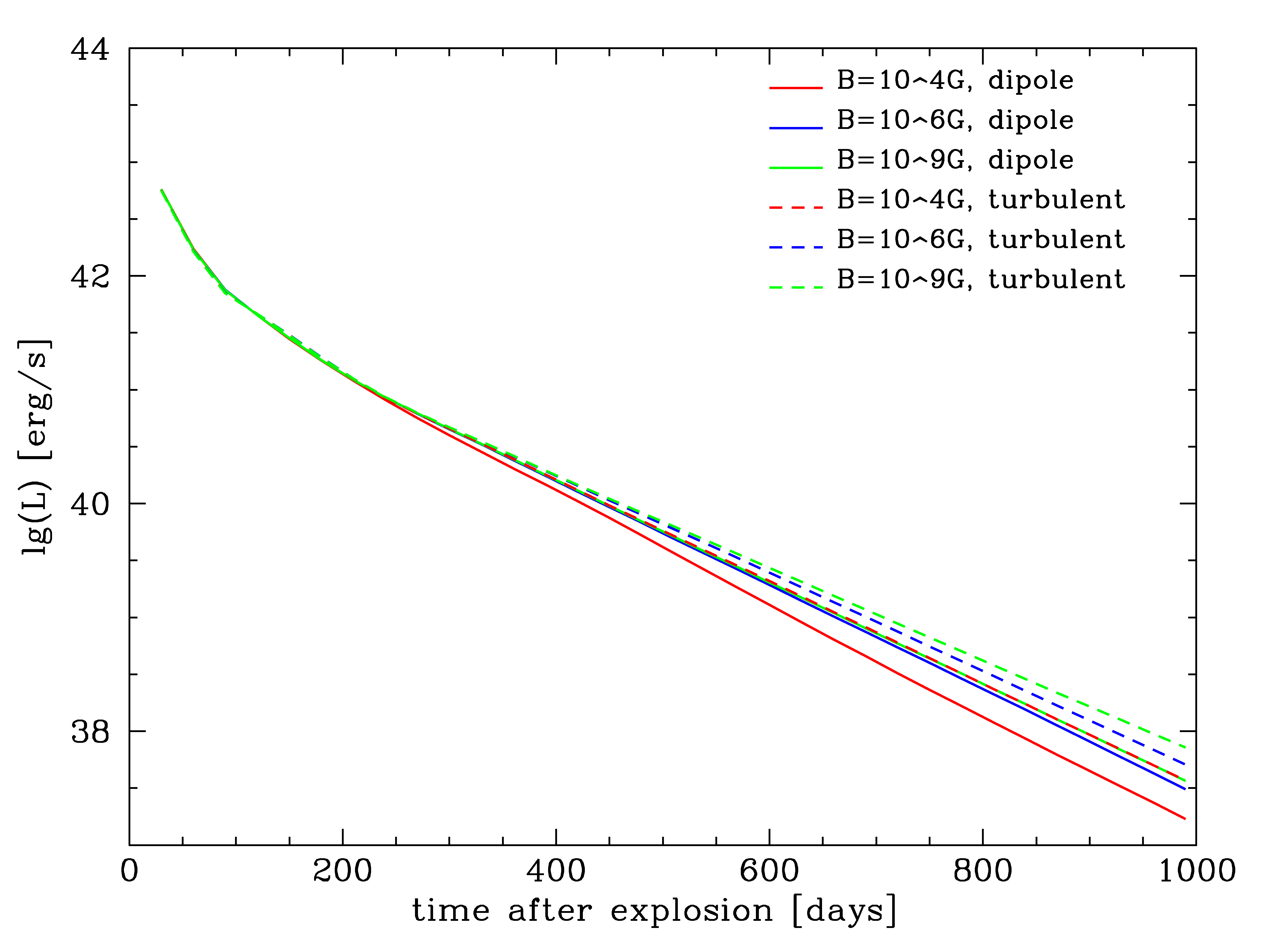} 
\includegraphics[width=0.44\textwidth]{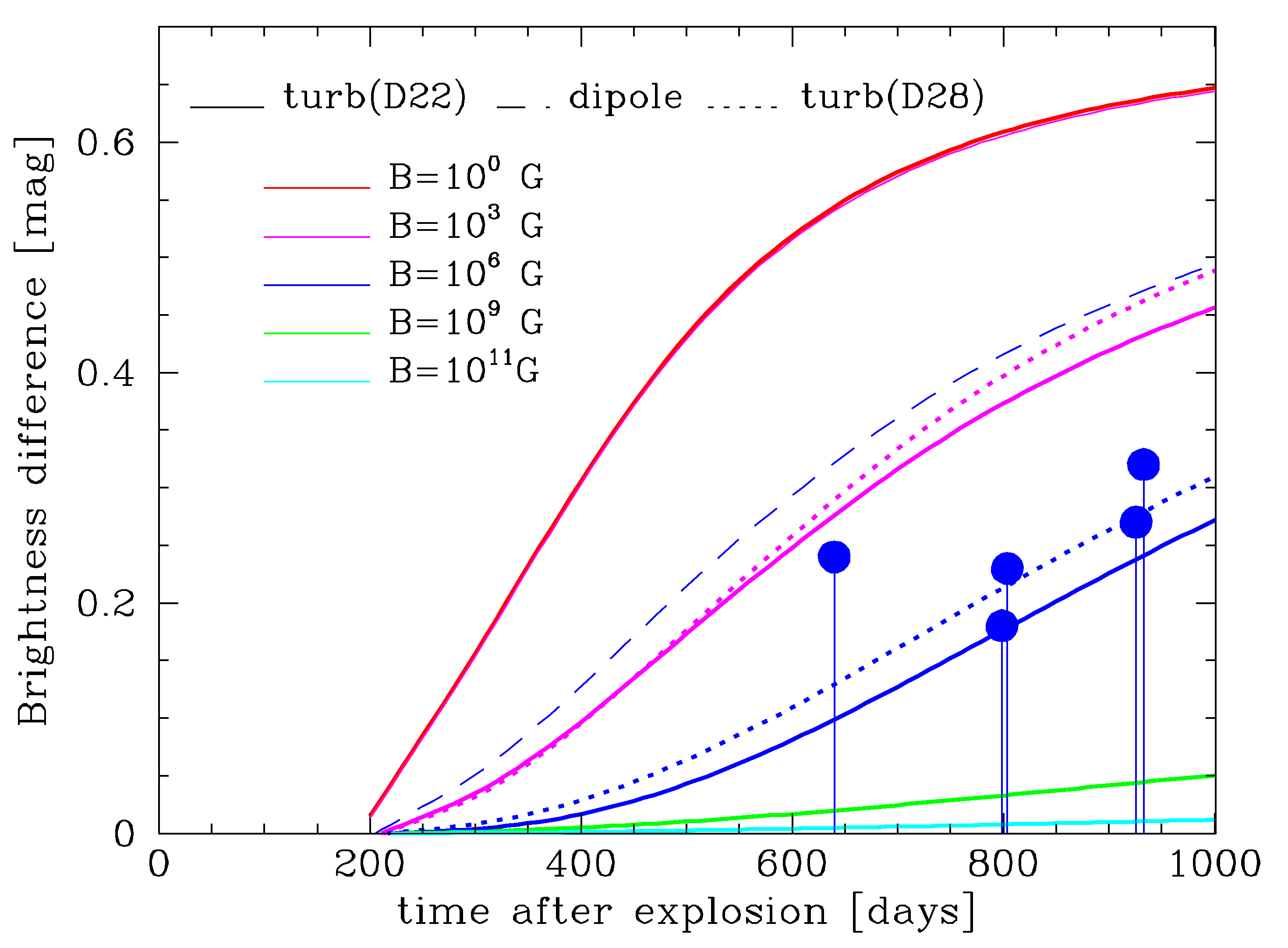} 
\vskip -0.0cm
\caption{{\bf Light Curves.} \newtextb{ Influence of the magnetic field morphology on light curves for
large-scale undisturbed dipoles (\emph{dipole}) and the field structure produced
by the turbulent morphology produced by Rayleigh-Taylor
instabilities starting from and initial dipole field in the WD for model D22 (solid line) and D28 (dotted). The left plots shows the light curve for a fixed dipole that does not evolve (solid) and the turbulent field taken from D22.  The right plot shows the evolution of the deviation with respect to our fiducial light curve to show the differences between D22 (solid line), D28 (dotted line) and the fixed dipole (dashed).}
As
background explosion model, we use the delayed detonation model 23 with parameters from
\citet{Hoeflich2017} for a normal-bright Type Ia Supernovae, and imprinted the
B-morphology with surface fields up to $10^{13} G$ which causes local trapping of positrons (see Table 2).  We show the
corresponding bolometric \acfp{LC} (left), and the difference (right) relative to the case of full local trapping. Note that a field size of ${10}^{11} G$ is already close to local trapping even at 1000 days. The details of the morphology of a turbulent field has minor impact (D22 vs. D28), 
but large scale ordered fields allow
\newtextb{for a significantly larger fraction of positrons to escape.}
As a result, the solution for the dipole field of
$10^3G$ is almost indistinguishable from the field-free case and its field of $10^6G$ produces an escape similar to a turbulent field but some three orders of magnitude smaller.  In addition, we give the loci of the upper limits for SN2012cg, SN2003hv, SN2014J, SN1992A and SN2011fe (blue dots from left to right). All SNe are consistent with \newtextb{complete trapping of positrons.}
Note that an assumed uncertainty of $0.15$ to $0.2^m$ for bolometric \ac{LC} implies the need for observations beyond $\approx 500 days$. As discussed in the text, all \ac{LC}
observations follow the case of local trapping of positrons (green dotted). At about 700 to 1000 days, additional radioactive heating places their locus above the full trapping case limiting the use of LCs to times before day 1000. For the estimates of the lower limit in $B$ (Table 3) we assume data starting beyond $\approx 250 $ days using $L_{bol}$ for a normal bright SNe~Ia (Paper II). Note beyond day 300 more than 95\% of the
energy input is from positron (see text).}
\label{LC} \end{center} \end{figure}

\section{Stage 2 Evolution: Observable consequences and implications of late time evolution}
\label{sec.stage2}

Based on the results described in the previous section, we will show the \ac{MHD} effects discussed above 
 based on the imprint on the late-time light curves \citep{Milne2001,penney14} and \ac{NIR}
 line profiles \citep{h04,penney14,tiara15} in the framework of delayed-detonation models \citep{khok89,hk96}.
 
In the stage 1 simulations, we studied the evolution of the morphology of 
the $B$-field based on the initial density structure and the adiabatic coefficient
$\gamma$ consistent with the equation of state used in the explosion simulations
of \hydra, namely the parameters of the spherical delayed-detonation model 23 of
\cite{Hoeflich2017}. 
In stage 1, the size of the B is a free parameter. 
This model starts with a WD originating from a main sequence
star of $7~ M_\odot$ and solar metallicity $Z_o$. The accretion rate from the
donor star has been tuned to produce a thermonuclear runaway at a central
density $\rho_c = 2 \times 10^9 g/cm^3$. 
In stage 2, we followed the
deflagration phase during the distributed regime of burning and the subsequent
detonation phase. The delayed-detonation transition has at a transition density
$\rho_{tr} = 2.3~10^7 ~g/cm^3$. We
triggered the delayed-detonation transition 'by hand' because the mechanism(s)
leading to the DDT have not been established. We choose this model because it can reproduce the
observed  curves, color-magnitude relation,  optical to mid-IR spectra of a
typical normal-bright SNe~Ia, namely SN2014J
\citep{tiara15,Telesco2015,Hoeflich2017}.

As in previous studies on the effects of $B$ on {positron} and photon
transport, we consider the nebular phase when the optical depth is and densities
are low, wherein forbidden atomic transitions dominate. The instant energy input
from radiative decay of $^{56}Co$ governs the bolometric luminosity. Moreover,
during the late phase, the envelope is homologeously expanding, so that position
and velocity of a mass element are related as $ r(m,t) \propto
|\vec{v}(m,x,y,z)| \times t = v(m) \times t$. Only the radial component of
$\vec{v}$ remains. The line profile of unblended features is a result of the
Doppler shift of the mass element $m$ and the energy deposition.

The \hydra ~code employs  full-3D
transport and hydrodynamics including the time dependence (see Appendix
\ref{appx:hydra}).  However, the results
in the nebular phase are dominated by forbidden lines in an mostly optically
thin envelope. The solution of the statistical equations is dominated by the
branching ratio between atomic energy levels $j$ and $i$, here determined
predominately by the Einstein values $A_{ji}$ \citep{Motohara06,penney14,tiara15,Telesco2015}, and the radiation transport is close to the optically thin limit leading to the stability of the results.

As discussed below, the \emph{lower} limits given below for the size of $B$-field derived from spherical LCs is not affected by $^{56}Ni$ plumes or instabilities.   This is because
mixing out of $^{56} Ni$ would
increase the escape probability of positrons and $\gamma$-rays  (see Fig.
\ref{LC}) which would increase the estimate for $B$, but our lower limits will still be valid.  The lower limits correspond to average surface fields assuming a magnetic topology identical to $D22$, for which our simulations of stage 1 show small deviations from sphericity. Even using the slightly less spherical model $D28$ results in only a slightly higher field estimate, see Figure \ref{LC}.

Note that we make the case that asphericities in $\rho $ for the highest $B$-fields may not survive the further evolution based on pure 3D hydro-simulations \citep{k01,g03,g05,roepke12}. However, this has not been shown in this work. 

We use the morphology of the magnetic field for model D22 at 0.6 seconds, and scale the average surface magnetic field between 1 and $10^{13} G$. For a grid of magnetic fields 
of $10^{0,1,3,4,6,9,11,13}$ G, the positron, $\gamma$,  and photon transport problem has been simulated at 100, 150, 200, 300, 400, 500, 750 and 1000 days.  For verification of our main conclusion, we have used the imprint of the highest magnetic field hydro-model (D28) with $10^{13}G$. We note that  large scale asymmetries in the density and chemical structure will affect the line profiles, as shown in previous simulations for 3D turbulent or off-center delayed-detonation models  \citep{2002NewAR..46..475H,hoeflich2006b,Motohara06}.

\subsection{Magnetic field effects on late time light curves}
\label{evolution2}
The results can be well understood within the following framework. $B$-fields
are frozen in comoving frame because of the plasma. Positrons are scattered and
lose energy by collisions with electrons till the energy is below the binding energy. At this point, they are annihilated. For details,  see appendix B.

To first order, the SN envelope
is freely expanding. The density drops with time $t$ as ${1/t^3}$ and the length scale increase $\propto t$. As a result, the mean free path of positrons \newtextc{relative to the expanding envelope} increases
like $t^2$ before annihilation.

In \newtextc{the} presence of $B$-fields, positrons gyro around the field lines with the
Larmor radius. If the structures in the field morphology are smaller than the
Larmor radius, positrons are trapped. Because structures grow linearly with $t$
but the Larmor radius increases like $t^2$, positrons will eventually propagate beyond these structure (see Table \ref{larmor}).
\begin{table}
\begin{center}
\caption{Maximum Larmor radius as a fraction of the size of the supernova envelope at various times  and initial surface field strengths.  During the nebular phase, the expansion is homologous with the distance of a mass element $r(m) \propto v(m) \times t $. $v(m) $ and $r(m) $ are equivalent measures of the envelope structure. However, $v(m)$ corresponds to a spectral  Doppler width observable in spectral line profiles.
Here, the size of the envelope is defined by the layer expanding with a velocity of 40,000 km/sec. In our simulations, the typical size of the $B$-field structure is produced by \ac{RT} instabilities ($\approx 500 ... 1,000 km/sec$ in velocity units or $1...2\times 10^{-2}$  in units of the envelope size.)}
\begin{tabular}{|c|c|c|c|c|c|}
\hline field/time & 100d & 300d & 500 d & 750 d & 1000d \\
\hline $10^3 G$ &  1.9 & 5.7 & 9.6 & 14.4 & 19.1 \\
\hline $10^4$G &  .19 & .57 & .96  & 1.4  & 1.9  \\
\hline $10^6$G &  .001 & .005 & .009 & 0.013 &  0.02 \\
\hline $10^9$G &  1.9E-6 & 5.E-6 & 9.E-6 & 1.3E-5 & 2.E-5 \\
\hline $10^{13}$G & 5.E-11 & 1.2E-10 & 2.E-10 & 3.3E-10 & 5.E-10 \\
\hline
\end{tabular}
\label{larmor}
\end{center}
\end{table}

Theoretical bolometric \acp{LC}, $L_{bol}$, and the difference in magnitudes in $V$,
a proxy for $L_{bol}$, are shown in Fig. \ref{LC} \newtext{for magnetic fields between 0
and $10^9...^{11}$G. The zero-field
case is identical to the $B=10^{3...4}\rm{G}$ dipole field, and the total local trapping case is virtually identical to the  $B\geq 10^{9}\rm{G}$ turbulent cases for all times.
The early \acp{LC} are identical because $\gamma-$ray transport does not depend on $B$ and
the free mean path of positrons is small.  At about 220 days, the energy input
from positrons equals that from $\gamma $-rays. Light curves with different
magnetic fields start to diverge
after about day 300 when the mean free path some of the positrons becomes large compared to the envelope.
Small scale structures and high $B$-fields trap positrons for longer than low
$B$ and or large scale dipoles. \newtextb{The luminosity for the simulation with  $B=10^9G$ and a turbulent field
follows closely the energy input from the decay of $^{56}\rm{Co}$ even at day 1000, 
which shows that the positrons are not able to 
escape the region of their birth.}  The difference (Figure
\ref{LC}, {right}) between the models may be in excess of $0.5 ^m$ which 
should be easily measurable given current observational accuracy.  We examine
observed \acp{LC} in the next section.}

\subsubsection{Magnetic field strengths in known supernovae}
   We want to combine the results of Figure \ref{LC} with observations.  
 Using our simulations (Fig. \ref{LC}, {right} plot) and applying the deviation
 of the observed range in time, a given \ac{SNeIa} provides estimates for the size of
 the $B$-field.
   The
   detailed reconstruction of $L_{bol}$ is beyond the scope of this paper and,
   thus, we refer to $L_{bol}$ (or $V$ as proxy) given in literature. 
   
  We use the radioactive tail of the fiducial light curve that keeps the
  positrons local, and compare it to simulations with a variety of magnetic
  field strengths and topologies. We then compare these differentials to observed
  supernovae to determine lower limits of their field strengths. {This can be
  seen in the right panel of Figure \ref{LC}, which shows the deviations from
  full trapping for each of our model simulations, as well as the upper limits on
  the deviation (thus lower limits on the field strength) for several known
  supernovae.  }

  Relatively
  few late-time \acp{LC} are available. \citet{Milne2001} suggested some indication
  for positron escape by day $\approx 170$,  but this was based on strictly
  radial $B$-fields in the models, and neglected near and mid-IR emission.
  Multi-band \acp{LC} are required to reconstruct $L_{bol}$ with sufficient
  accuracy to include the redistribution of photons to the \ac{NIR} and \ac{MIR} within
  an accuracy of $\approx 0.15^m$ \citep{stritzinger02,Sollerman2004,Gall2018}.
  If \ac{NIR} data are not available, we use the
  redistribution functions from our models for normal bright and transitional
  \ac{SNe} \citep{Hoeflich2017,Gall2018} and redistribution functions based on observations of SN~2014J
  \citep{Telesco2015}. We use the intrinsic uncertainty from the scatter in the
  measurements. \newtextb{For $L_{bol}$, based on the references above,
  we add to the measurement error in the LC points an additional error of $0.15^m$  to the reconstruction of
  $L_{bol}$ if \ac{IR}  observations are available, and $0.2^m$ if they are
  not.}

\begin{table}
\centering
{\bf
\caption{Lower limits for magnetic field strengths from known supernovae. The dates correspond to the last data point observed.
    The loci in Fig. 11 are the last data point before additional energy sources contribute to the light curve.
}
\label{table.prediction}
\begin{tabular}{|l| l| l| l| l|}
\hline
        Name & t [days] &  $B_{\rm{turb}}$ & $B_{\rm{dipole}}$ &reference\\
        \hline
        SN1992A   & $926$  & $10^{5.5}$ & $10^{8.5}$ &\cite{Cappellaro1997}\\
        SN2003hv  & $800$  & $10^6$ & $10^9$  &\cite{Leloudas09} \\
        SN2011fe  & $930$  & $10^5$ & $10^8$&\cite{Kerzendorf2014}\\
        SN2012cg  & $640$ & $10^4$ & $10^7$  & \cite{Graur2017} \\
        SN2014J   & $800$ & $10^{5.5}$ & $10^{8.5}$ &  \cite{yang2018}\\
        \hline
\end{tabular}
}
\end{table}

  \newtextb{None of the observations fall below the light curve expected from  the decay
  of $^{56}\rm{Co}$, indicating full local trapping by magnetic structures.}
  To determine the lower limits of the initial $B$,
  we use the curves of Figure \ref{LC} and determine the maximum deviation from the theoretical 
  curves which are consistent with the decay of 
  $^{56}\rm{Co}$. Obviously, this limits depend on the coverage of the late-time LCs and the error bars.
  Observed \acp{LC} include SN1992A ($\approx 950d$, \cite{Cappellaro1997}), SN2003hv
  ($\approx 700d$, \cite{Graur2017}), SN2011fe ($\approx
  930d$, \cite{Kerzendorf2014}),  and SN2012cg ($\approx 1060d$, 
  \cite{Graur2017}).  {SN2014J} has been observed even beyond 1200 days, showing
  the onset of radioactive decay of elements besides $^{56}\rm{Co}$. This
  flattens the \ac{LC} by about day 800, indicating the onset of $^{57}Co$ decay
  as \newtextc{the} dominant energy source \citep{yang2018}. 
  SN2014J was found to be consistent $^{56}\rm{Co}$ and with no signature for positron escape nor an \ac{IR} catastrophe predicted by  \cite{fransonn94}. 
  Similarly, SN2003hv, SN2011fe
  are observed after onset of $^{57}\rm{Co}$ heating or interaction, and
  neither LC drops below the $^{56}\rm{Co}$ or $^{57}\rm{Co}$ line, giving
  evidence for full positron trapping.
  Results for the lower limit are summarized in Table \ref{table.prediction}.
This shows that high $B$-fields are common, well in excess of typical fields observed in \acp{WD} (see Sect. 1).  
Note that, in the presence of a deflagration phase in \myMCh-mass explosions, large
scale dipole fields will be transformed into small-scales turbulent fields.  Alternative explosion scenarios involve   pure detonation burning. For these, the  
 dipole field morphology may remain (see next section) leading to even larger lower limits for the initial $B$ field.

\begin{figure}[h]
\centering
\includegraphics[width=0.69\textwidth]{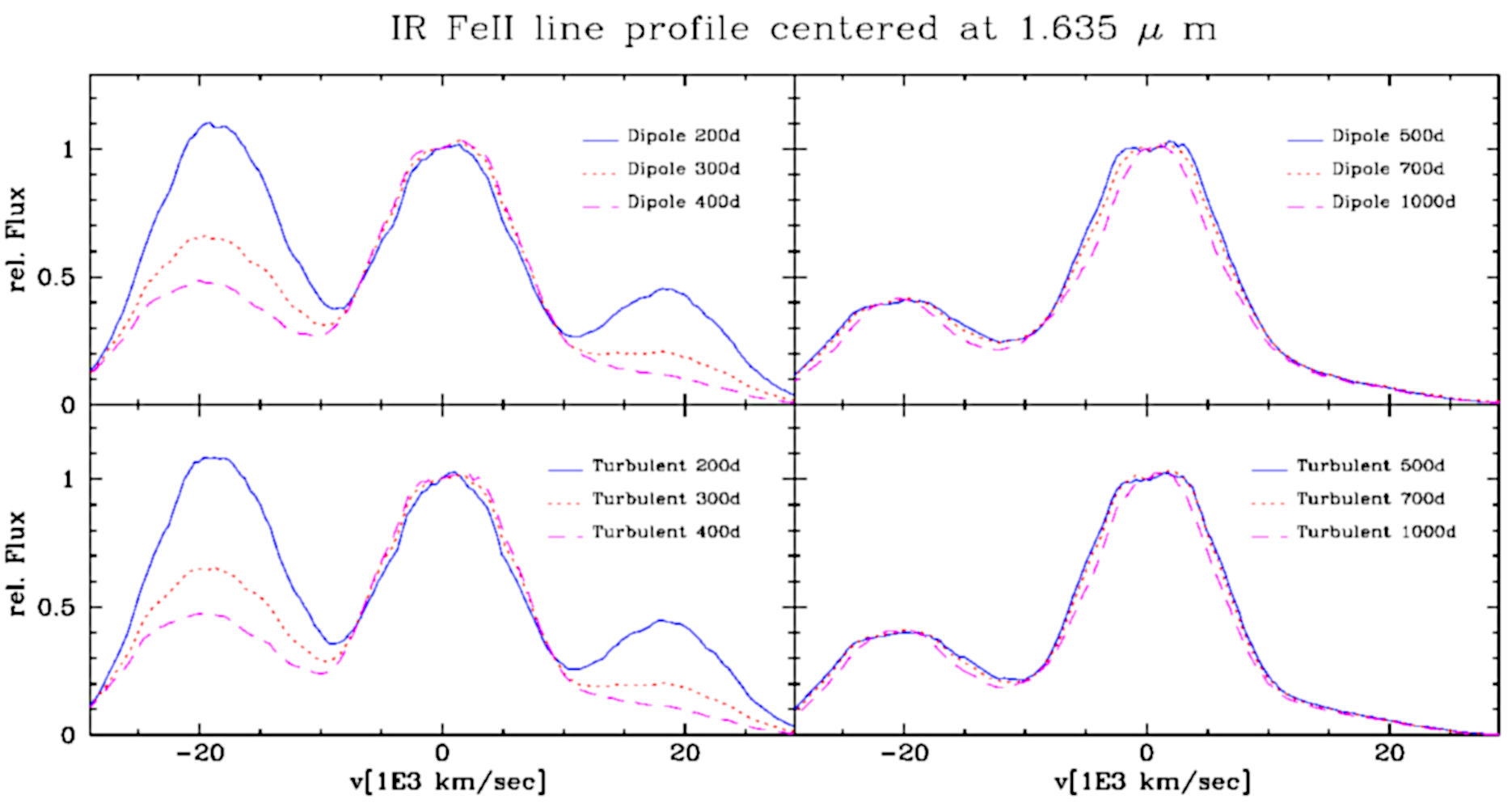}\\
\vskip -0.1cm
\caption{ {\bf Angle averaged line profiles} for the 'almost' unblended feature
produced by the forbidden [Fe II] at $1.644 \mu m$ for an initial average surface field of $10^6G$.
The time evolution for the dipole (top row) and turbulent field (bottom row) shows little evolution before 400 days (left column) due to the local trapping of positrons.  
For dipole and turbulent fields, the profiles start to change
after about 400 and 700 days, respectively (right column). 
A total of $\approx$ 100,000 line profiles have been calculated (magnetic field*morphology*phases*$\theta$ and $\phi$ direction $ \approx 10^5$), are available on request, and will be part of a data base for different explosion models.}
\label{fig12}
\end{figure}

\begin{figure}[h]
\includegraphics[width=0.49\textwidth]{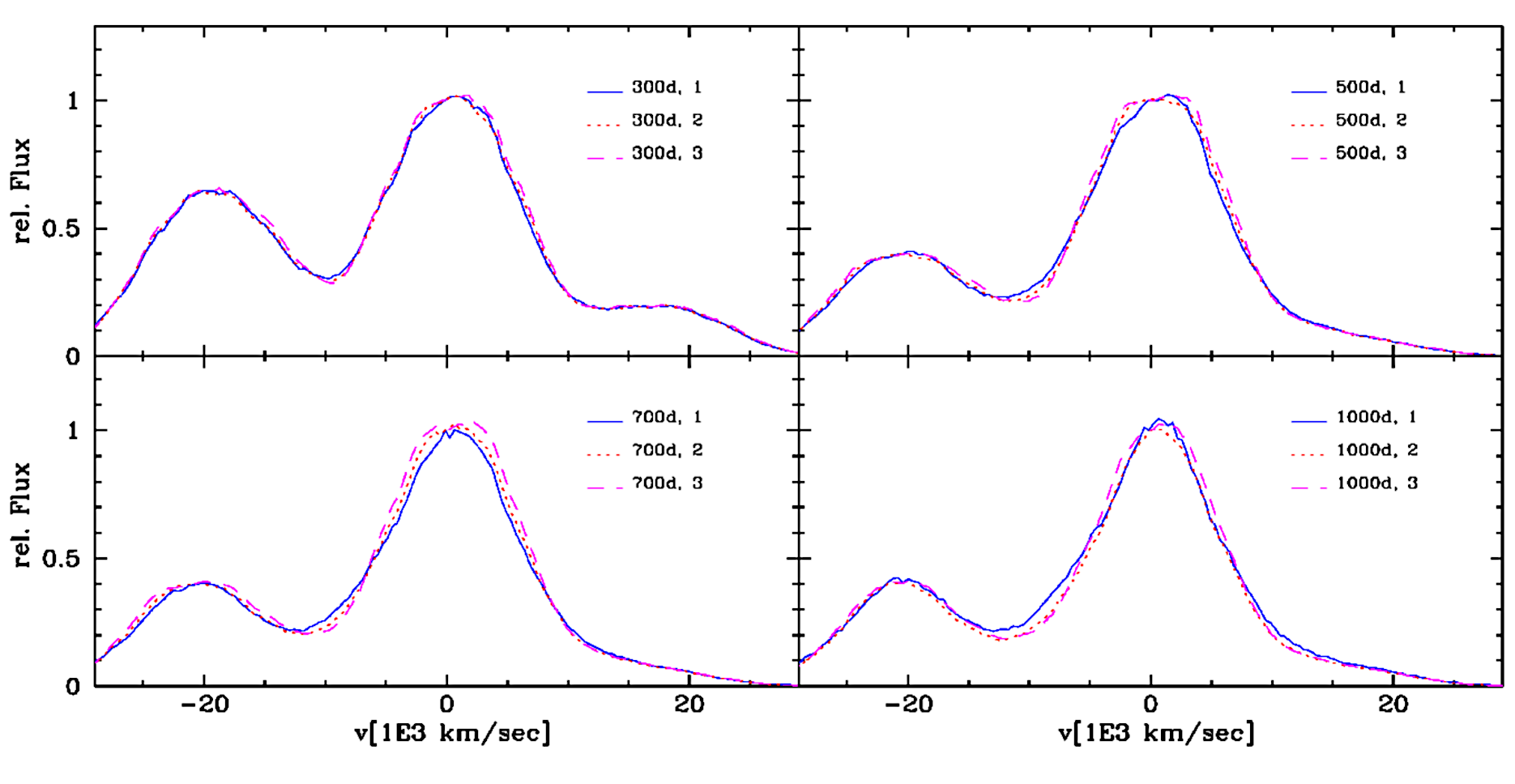}
\includegraphics[width=0.49\textwidth]{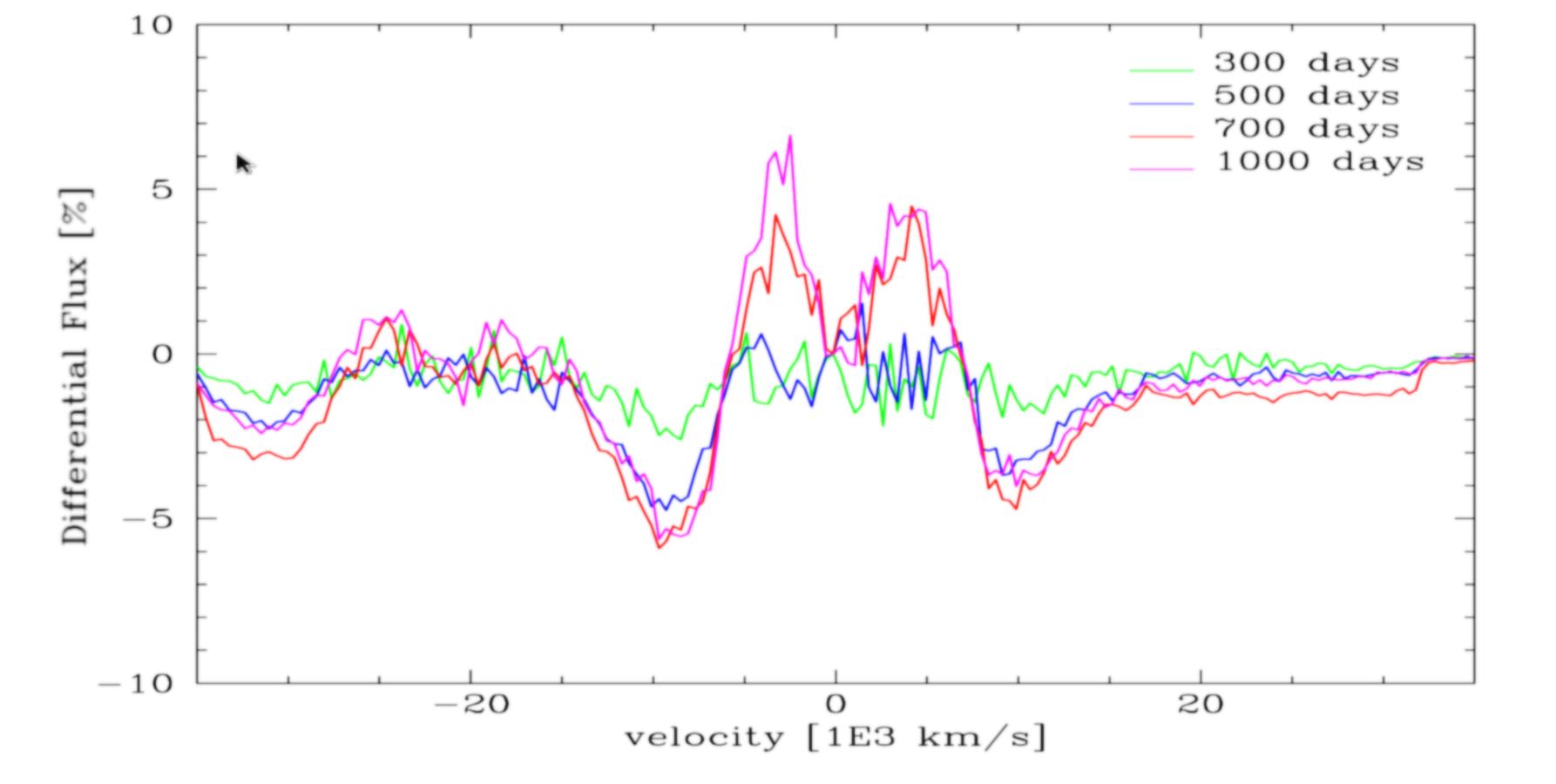}
\vskip -0.1cm 
\caption{
\myedit{Same as Figure 
\ref{fig12} 
but
[Fe II] line profiles for the dipole field 
as seen from $0^o$ (\emph{1}), $30^o$ (\emph{2}) or from the pole (\emph{3}) at given times (left). The late-time differences in the profiles ($\sim 1,000 ... 15,000~ km/s$), although not prominent on this plot, are significant, given a typical spectroscopy resolution of $\approx 100 ~km/s$. In addition and as proof of concept, differences between NIR spectra between the dipole seen equator on and turbulent field are shown on the right with spectra normalized to the average spectra at day 500. Dipole fields produce a characteristic, large scale pattern and signals of about $\pm  5 \%$ between $\pm 15,000 km/sec$. For turbulent fields, variations correspond to the mass of the individual plumes and show up as wiggles on the 1\% level. Note that spectroscopy allows precise measurements of Doppler shifts but flux differences on the 1\% level will be reserved to observations with the upcoming JWST. For details, see text.
}
}  
 \label{IR-spectra-angle}
\end{figure}

\subsubsection{Magnetic field effects on line profiles}

\acp{LC} are a powerful tool but, their application is limited as tools to constrain $B$ because \acp{LC} tend to flatten after 700
to 1000 days.
  Line profiles and their variations as a function of time have been suggested
  as an alternative diagnostic tool in Paper II.

  In particular,  [FeII] at $1.644 \mu m$
  has been identified having only minor blends in the line wing
  \citep{h04,Motohara06,penney14,tiara15,Diamond2018}. All other strong features
  at shorter wavelengths are blends produced by multiple transitions.
  Alternatively to the NIR [FeII], the \ac{MIR} has been demonstrated as source
  for unblended [Co II] and [FeII] lines by \citep{Telesco2015} but, likely,
  those are beyond reach but for the upcoming JWST.  Here, we want to discuss the
  \ac{NIR} [Fe II] (Figures \ref{fig12} and \ref{IR-spectra-angle}). 

  The profile is a measure of the
  distribution of the energy input by positrons convolved with the Fe abundance.
  The line profiles are more sensitive  to positron transport effects because
  changes do not require the escape of positrons, but non-locality of the
  positron decay is sufficient \citep{penney14}. It provides information
  about the distribution and possible off-set of radioactive $^{56}Ni
  \rightarrow ^{56}Co \rightarrow ^{56}Fe$ and, at ultra-late times, other decay
  channels of radioactive isotopes of the iron group \citep{Hoeflich_Prog2019}.
  For the time series, we use the small variation in the line profile as
  indicator for $B$-fields.  

   Inherently, the effects of $B$ can be detected
  earlier, namely by day 400...500 and, in principle, without restrictions to
  late times well  beyond 1000 days.  However, optical depth effects, namely the
  underlying optically thick photosphere  and non-local energy input by
  $\gamma-$ rays for iron close to the center, leads {to} a
   change of the forbidden [Fe II] {line} profile at
  1.644 $\mu m$, rather than revealing positron transport effects (Fig.
  \ref{fig12}, left). However, the {line} profile  hardly changes between 300 and 400 days. This requires a reference point
  for the spectra later than 300 days.  Note  that the two features to the right
  and left correspond to [Co III] and [FeII] blends and variations that are caused by
  the nuclear decay with time \citep{Hoefich04}. In practice, one
  problem is related to the costs of obtaining \ac{NIR} spectra of high quality, and the
  need for a time series, which limits the number of objects to a few (see  Sect.
  1).

 As an example and for a field of $B=10^6G$, the evolution of the profiles is shown
 as a function of time and orientation in Fig. \ref{fig12}. For both
 turbulent and dipole fields, the [FeII] {profiles remain almost unchanged} \newtextb{between 300 and
 400 days after the explosion, because $\gamma-$rays do not contribute substantially to the energy input, and positrons annihilate locally.}
 After about day 500, the profile becomes
 increasingly narrow because positrons at the outer, high-velocity, low-density
 regions of the $^{56}Ni$ distribution become non-local, \newtextc{that is they move
away from their generation region and excite other lines and elements
within the envelope}.

One fundamental problem to decipher the distribution of the radioactive
$^{56}Ni$ is that  a large scale  $B$ field, e.g. a dipole,  causes strong directional dependence  when a particular \ac{SNeIa} is
observed.  The dependence of the line profile on viewing angle can be seen in
Figure \ref{IR-spectra-angle}.  
In our example of a large-scale dipole, positrons can travel almost unhindered
along the axis of symmetry, the polar direction, {thus the effective mean-free path has a small dependence on $B$} as opposed to the
highly inhibited travel along the equator. The changes between 500 and 750d and
500 to 1000d are comparable when seen from the equator and pole, respectively.
The rate of change does depend on both the size of $B$ and the direction of the
observer.  
   
For the turbulent field, the profiles show little directional dependence.
Fluctuations are restricted to the scale of the convective eddies which, in
principle, may be recovered by high signal-to-noise observations with a
resolution of the characteristic eddy-velocities, i.e. 1000 km/sec. We note that
the evolution of the profile, e.g. the mean-half width is non-monotonic
\citep{tiara15} and, thus, requires at least 3 observations to recover.
    
One of the main result of our \ac{MHD} simulations is that for a wide range of $B$ the morphology is given by 
a turbulent field even for a large-scale initial dipole during the deflagration phase. 
This does not rule out large dipole fields in, e.g., pure detonation scenarios.

  \section {Final Discussion and Conclusions}
  \label{sec.discussion}

\newtext{
We have studied the effects of magnetic fields on deflagration fronts and
late-time spectra in Type Ia supernovae.  We find that small scale magnetic
fields develop as a result of the motions of the burning front, \newtextc{and that
observations put the magnitude of  magnetic fields into a regime which may be
both relevant and larger than commonly observed in single WDs.}}
  
 We employed initial maximum fields between $B=10^7\rm{G}$ and $10^{15}\rm{G}$, 
\newtextc{in} the
 framework of delayed detonation models.
We examine their imprint on light curves and spectral properties during the nebular phase.
Using our model light curves, we estimate lower
limits of $10^6\rm{G}$ for several published supernovae.  We also find that
Rayleigh Taylor instabilities develop early, even in the presence of magnetic
fields that, energetically, should suppress the instability.

The two weakly magnetized runs, D22 and D26, are almost entirely identical in
their evolution in all measurable quantities.  This is unsurprising, as the
magnetic pressure is several orders of magnitude lower than the gas pressure at
all radii for both of these simulations.  This also shows that we are not
troubled by systematic variations other than those caused by the magnetic field.
Tangling develops early (0.5 s) in all runs. There is slightly less tangling in
the strongly magnetized simulations, as well as more pronounced \ac{RT}
instability.  The strong initial field in D28 supports large-scale flow
structures, which results in faster increase in rising plumes as well as
reduced burning surface area. This can explain why D28 has 15\% less \myNi~ vs.
lower $B$ as in \citet{Hristov2018}.

We do not see any evidence of magnetic field amplification in these simulations
as expected from previous studies \citep{Hristov2018},
though this does not rule out amplification with other initial magnetic topologies.

We
started our simulations with a burned central region which, in size, corresponds
to about 1 to 2 seconds after the thermonuclear runaway in models of central
ignition.
This is the onset of RT in models with central ignition \citep{khok03,gamezo04}. 
In off-center and multi-spot ignitions, 
the development of \ac{RT} instabilities is faster \citep{hkrr13}.
Though we do not follow the deflagration front during the regime of distributed
burning, the tangling cannot be expected to decrease. Rather, in the absence of
a large scale flow to order the field, it can only be
expected to increase during the regime of distributed and detonation burning.
We show that the density
distribution remains spherical to within $<1\%$ in the most strongly magnetized
case. This is due to the dominance of pressure
equilibrium and gravity, a result found previously by hydro-simulations for
\ac{QSE} of explosive oxygen burning regions \citep{g03,g05}.

\newtext{
 In Section \ref{sec.stage2}, light curves between 500 and  1000 days are
shown to be able to constrain the strength of the magnetic field, and we use
these model light curves to estimate magnetic field strengths of $>10^6\rm{G}$
from several observed supernovae.
Line profiles, or rather their lack of time-evolution, can be
used to constrain $B$ starting \newtextc{at} about day 300. For
 magnetic fields larger than $10^6\rm{G}$, positron are trapped within
the magnetic field up to about day 500 which presents the current limit from observations. However, changes in the profiles or the width of
the line can be used to determine the magnetic field strength for larger initial fields because, 
in principle, the method of using
specific line profiles can be applied to even later phases than light curves.
Currently, only a few \ac{SNeIa}
with sufficiently late time \ac{NIR} spectra or with optical and \ac{IR} light
curves  are available.  A larger number is needed to confirm whether high $B$
fields are generally present. Figure \ref{LC} may provide a tool to get a
$1^{st}$ order estimate $B$ fields of future observations.
}

\newtext{
As mentioned in the introduction, an obvious question is the origin of such a
field.  The magnetic decay timescale of $10^9\rm{K}$ plasma is $10^9\rm{yr}$, so
it is natural for such a plasma to support magnetic fields, and the question is
identifying its source.  Within dynamo theory \citep[e.g.][]{Brandenburg05} 
there are several mechanisms to amplify the field up to the saturation strength
of $\sim 10^{14}\rm{G}$
\citep{Chandrasekhar56a,Chandrasekhar56b,Mestel56,Hristov2018},
 and the nature of the amplification mechanism will be
imprinted in the magnetic field structures.    This may happen at several points during the evolution
of the WD.  In the Chandrasekhar mass, \myMCh, explosions considered here, a WD
close to equilibrium begins to burn as a result of compressional heating, which
in turn results from accretion from a companion.  This leads to subsonic
deflagration, which then transitions to supersonic detonation.  The correlation length
of the flow during each phase will set the correlation length of the magnetic field, which in
turn impacts the escape of positrons.  During the accretion phase, the dominant
length scale is the radius of the white dwarf.  During the late-stage run up to
the deflagration, the so-called \emph{smoldering phase} \citep{hs02,Zingale11}
the dominant length scale is the pressure scale-height of the star.  During the
supersonic explosion, the scale is set by the sound-crossing time scale, \newtextb{because the flame propagates as a weak detonation.}
  Our results
show that amplification during the deflagration phase is unlikely.
Hydrodynamical simulations \citep{g03,roepke12}
have shown that the instabilities that could give rise to a dynamo are frozen out during the expansion phase of
the WD.  This leaves the accretion phase or smoldering phase as the likely
candidates for magnetic field amplification in WDs.
}

\newtextc{  This leads us to 
  the discussion of these results in the framework of alternative explosion
  scenarios.}
Likely, a combination of different explosion scenarios are realized in nature.
The quest for the predominant
mechanism is still under debate with time changing favorites. This is not too
surprising
, because it is nuclear physics that determines the
progenitor structure, \myeditt{the} explosion and \myeditt{the} subsequent \acp{LC} and spectra
\citep{hoeflich2013} and combined with an almost overlapping mass range for the progenitor \ac{WD}
among different scenarios. 
\myeditt{This phenomenon is also known as `stellar amnesia'. The mass ranges} for normal bright \ac{SNeIa}, \acp{HeD} are $\approx 1.0~
...~ 1.1 M_\odot $ \citep{pakamor12,Shen2018}, and 1.28 to 1.38 $M_\odot$ for \myMCh-mass
explosions \citep{HWT98,tiara15}.
\myeditt{For a longer review  of explosion scenarios, details and further references are given in the introduction, \citet{Hoeflich_Book} and, for individual aspects of the explosion, in the  
subsequent sections under the chapter titled "Physics of Thermonuclear
Supernovae" in \citet{2017hsn..book.....A}.}
  During dynamical merging of two WDs (e.g. \citep{Pakmor11}), high $B$ fields are likely to develop depending on details of the dynamical merging process. However, no significant $B$ field amplification can be expected for He-tiggered explosions because they explode promptly on time-scales of a second triggered by a supersonic shock, and the progenitor evolution involves two WD without a deflagration or smoldering phase. However, a large scale dipole  field may develop during the accretion phase similar to massive He/Co systems described in \citet{Pakmor21}.

Finally, we would like to discuss some limitations beyond those just mentioned which will be overcome in future. 
Our study is not a full end-to-end simulation of the explosion of a WD. Rather, we studied the change of the morphology of an initial magnetic field during the regime of non-distributed burning, and the observable consequences.
The initial condition of the 3D simulation starts from a WD structure but a rather advanced stage of burning to shorten the time till the RT-instabilities can develop. We do not follow the ignition process. At the end of the 3D simulations, the further evolution is followed by spherical radiation-hydro without magneto-hydrodynamical effects.

For our analysis of the \ac{NIR}-spectra and light curves we have considered a specific explosion model and realization. A more comprehensive study including spectra at 750d and beyond
is under way and encompassing transitional and sub-luminous \ac{SNeIa} and alternative explosion scenarios. 

\newtext{ \ac{NIR} spectra of SN2004du, 2005df and 2014J, and optical \acp{LC} of 5 SNeIa discussed in this paper suggest $B$ fields in excess of $10^{4...6} G$ to be common.
One obvious limitation is the lack of  observations for a larger sample.}
  With the upcoming James Webb Space Telescope, unblended line profiles of Co in the \ac{MIR} will become available
  (e.g. \citet{gerardy07,Telesco2015}),
  and WFIRST will cover the LC evolution of many local \ac{SNeIa} 'by chance'.

\acknowledgments
We thank the referee for carefully reading the manuscript and many helpful suggestions.
We acknowledge the support by National Science Foundation (NSF) grant AST-1715133 including  the salary for a postdoc. Most of the development and simulations have been done on the local cluster of the FSU astro-group including the data storage.
This work used the Extreme Science and
Engineering Discovery Environment (XSEDE, {\tt https://www.tacc.utexas.edu}, \citet{xsede}), which is
supported by the NSF grant ACI-1548562, under the 
XSEDE allocation TG-AST140008.

\bigskip 

\noindent
\emph{Software}:{ \enzo\ \citep{Bryan14}, \hydra\ \citep{h90,h93,2003ASPC..288..185H,2003ASPC..288..371H,2009AIPC.1171..161H,penney14,h17}.
Plotting and analysis was done using
\texttt{yt} \citep{yt}, 
\texttt{matplotlib} \citep{matplotlib}, and 
\texttt{numpy} \citep{numpy}.}

\bibliographystyle{apj.bst}
\bibliography{arxiv.bib,article.bib,article_add.bib,my.bib,article1.bib,article_mod.bib,articlesnew.bib,article_rev2.bib}

\begin{appendix}
\label{appx}

\section
{Mass and molar fractions and burning operator}
\label{appx:frac}

{
Here we define the mass and molar fractions in terms of other quantities
and provide some identities. Firstly note that in a $\myfuel$/$\myproduct$ mixture, 
the partial mass densities add up to the total mass density:
}
\begin{equation}
\label{eq:rhoparts}
\rho = \rho_{\myfuel} + \rho_{\myprod}
\end{equation}
\newtext{
Let the respective molar masses are $\myA{\myfuel}$ and $\myA{\myprod}$. The mass, molar, and burned fractions, $\myX{}$, $\myY{}$ and $f$, are defined as follows:
}%
\begin{align}
& \myX{\myfuel} \equiv \rho_{\myfuel} / \rho
,
\;\;\;\;
\myY{\myfuel} \equiv \myX{\myfuel} / \myA{\myfuel}
\\
& \myX{\myprod} \equiv \rho_{\myprod} / \rho
,
\;\;\;\;\
\myY{\myprod} \equiv \myX{\myprod} / \myA{\myprod}
\\
& f \equiv f_{\myprod} \equiv \frac{\myY{\myprod}}{\myY{\myfuel} + \myY{\myprod}}
\\
& f_{\myfuel} \equiv 1 - f
\end{align}

{
For the burned fraction $0 \leq f \leq 1$ always holds, 
whereas $f=0$ corresponds to pure fuel and $f=1$ to pure product. 
The two mass fractions obey $\myX{\myfuel} + \myX{\myprod} \equiv 1$.
}

{
Equation (\ref{eq:rhoparts}) allows for a code to keep track of one partial density variable in addition to the total mass density. In our case we had chosen the product density. The burning operator, eq. (\ref{eq:flame-diffusion}), is evaluated at the end of the time cycle. Below are the identities necessary to go back and forth between the relevant quantities:
}
\begin{align}
& f 
=
\frac{\myA{\myfuel}\;\rho_{\myprod}}
{\myA{\myprod}\;\rho + \left( \myA{\myfuel} - \myA{\myprod} \right) \rho_{\myprod}}
\\
& \rho_{\myprod} 
=
\frac { \myA{\myprod} \; f } 
{ \myA{\myfuel} (1-f) + \myA{\myprod} \; f}
\; \rho
\label{eqn.frac_prod}
\end{align}

{
All our MHD simulations assume the product being $^{56}$Ni, as well as $^{12}$C : $^{16}$O molar ratio of 50 : 50 for the fuel mixture. This corresponds to $\myA{\myfuel} : \myA{\myprod} = 1 : 4$.
Note that $\rho_{prod} $ is a tracer field not entering the MHD-equations.
}
\vfill\eject
\section
{Radiation Hydrodynamics  with \hydra}
\label{appx:hydra}
\begin{figure}[!b]
\epsscale{0.95}
\plotone{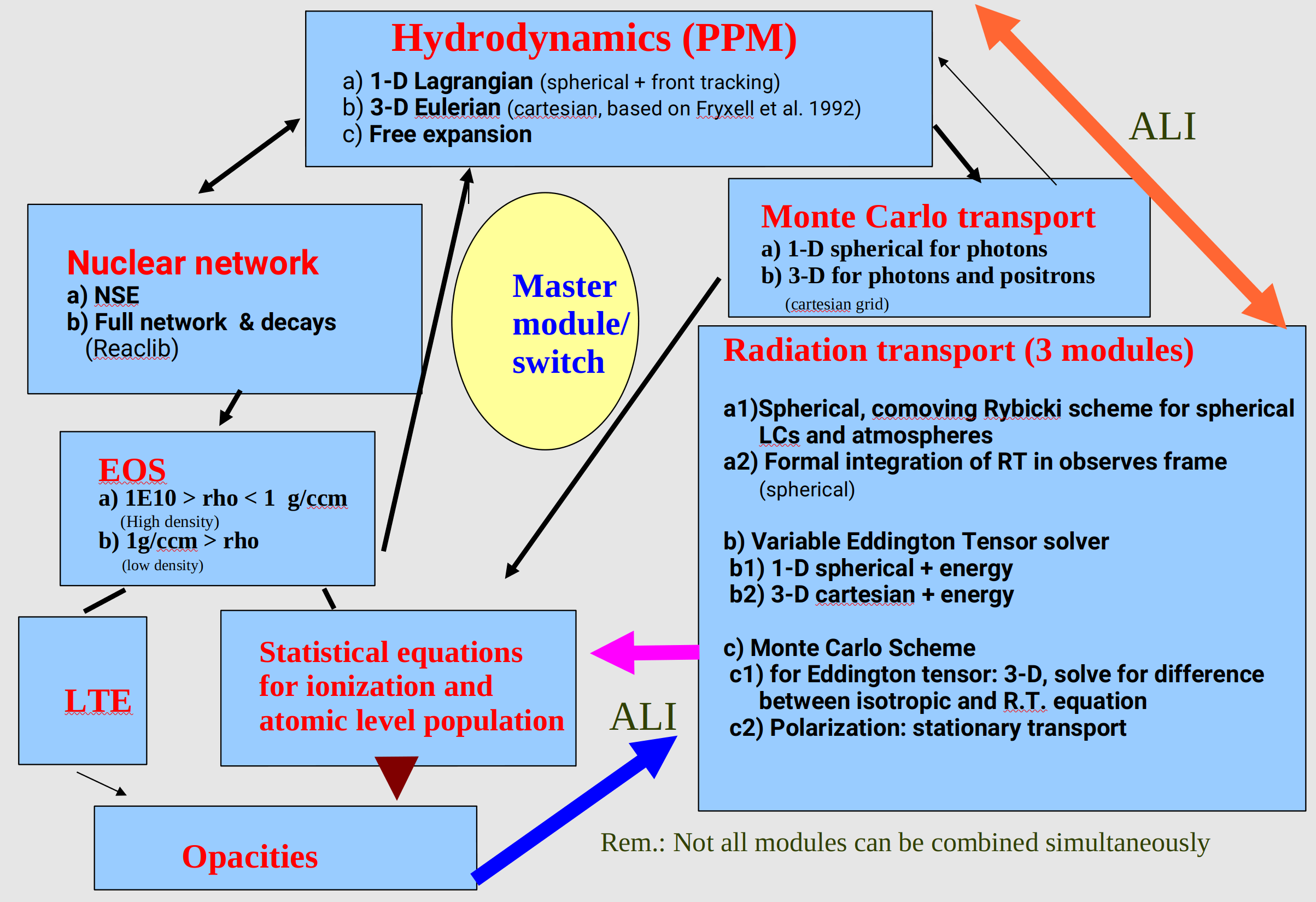}
\caption{
 Block diagram of our numerical scheme to solve
radiation hydrodynamical problems, including detailed equation of
state, nuclear and atomic networks. 
\dccq{Which of these modules are not used in This Paper?  Let's list those. Answer: All are used in this paper}}
\label{modules}
\end{figure}

Our HYDrodynamical RAdiation \textit{HYDRA} code consists of physics-based
modules to provide a solution for the 
nuclear networks, 
the statistical
equations needed to determine the atomic level population, 
the equations of state, 
the opacities, 
and the hydro  
and radiation problems. 
The individual
modules are coupled explicitly (Fig. \ref{modules}).  Consistency between the
solutions is achieved iteratively by perturbation methods, a combination of accelerate lambda acceleration plus equivalent-two level approach (ALI2) net-cooling and
heating rates for radiative coupling terms
\citep{2003ASPC..288..185H,2009AIPC.1171..161H}, and excitation and ionization by
hard radiation and particles.  
Different modules are employed during different
stages of the simulation. 
Below, we will give the basic equations with references to
the methods actually used in this paper, using
 standard notation as in
\citet{MM84_book} wherever possible.
Though several of the modules are
formally used as part of HYDRA but do not affect the results in this paper as discussed in the main text.
\dccq{Lets list those separately}

\noindent{\bf Hydrodynamics:} The structure of the expanding envelopes are
obtained by three different modules by a) assuming free expansion, or by solving
the non-relativistic hydro equations in b)  the Lagrangian frame for spherical
geometry including a front tracking scheme to resolve shock fronts
\citep{fryxell}, or c) the Eulerian scheme for full 3-D using cartesian coordinates based on PROMETHEUS \citep{Fryxell91}.{ 
In this paper, we use module b) for the early evolution till
$\approx day 10$, and module a) afterwards. }
The hydro modules use an
explicit Piecewise Parabolic Method (PPM) by \citet{CW_PPM84} to solve the
compressible reactive flow equations with variable adiabatic gradients based on
low and high density equation of state (see below).  PPM is implemented as a
step followed by separate remaps of the thermal and kinetic energy to avoid
numerical generation of spurious pressure disturbances during propagation of
reaction fronts (flames and detonations). \dccq{Those are the options, which was
used here?}

\noindent
{\bf Deflagration fronts:} For spherical geometry, 
the deflagration speed in mass coordinates is given by
\begin{equation}
v_t=max(v_{cond},C_1 \sqrt{\alpha_T g L_f}) 
\end{equation}
with $v_{cond}$ being the conduction speed, 
with the Atwood number $\alpha_T = (\alpha -1)/(\alpha +1)$ and  $\alpha
=\rho^+(r_{burn}) / \rho^-(r_{burn})$,  $r_{burn}$ being the distance of the front from the
center,  $\rho^+,\rho^-$ the density jump across the front.  { {$L_f$} is the characteristic length scale for the freeze out of the turbulence. The main effect of the expansion is the freeze out of the turbulence on scales $L_f$ where the turbulent velocity $v_t$ due to RT instabilities is comparable to the differential expansion velocities.} For these flows,  $v_t \approx v_{exp}=L_f~t$ with
$v_{exp}$ is the expansion velocity at time $t$.  $C_1$ has been calibrated
\citep{2000ApJ...528..854D}, based on 3D hydrodynamical simulations
\citep{khokhlov97,g03}. Here, we did use $C1=0.2$ . In other works where we employ 3D hydro, we use the
burning operator given in Equations 8 \& 9.

\noindent {\bf High-density equation of state and nuclear reactions:}
The EOS is for a partially degenerate and partially relativistic Bose- and
Fermi-gas plus interactions due to
effects of Coulomb corrections, quantum-relativistic effects on the electron
component, and electron-positron pair production
\citep{nomoto84,1931ApJ....74...81C,vanHorn69,slattery82,Goprdamp84,1931ApJ....74...81C,hoeflich06}. Electron screenings are taken from \citet{graboske73} and in the weak,
intermediate and intermediate-strong regime, and from \citet{itoh79} in the strong regime. 

\noindent{\bf Nuclear reactions} are used in the high density regime and
temperatures above $5 \times 10^5 K$.  A network is used with rates
including weak, strong and electromagnetic reactions. It is  based on the
implementation by \cite{T94a,T94b} but with modified matrix solvers. For isotope
$i$, the change of the abundance ratio per nucleon $Y_i$ is given by
\begin{equation} 
    \dot Y_i = \sum_j N^i_j \lambda_j Y_j + \sum_{j,k}
    {N^i_{j,k}\over 1+\delta_{jk}} \rho N_A \langle \sigma v \rangle_{j;k} Y_j
    Y_k + O(3). \end{equation} 
The first term on the right-hand side
includes single-particle processes;  decays, photo-disintegrations, electron and positron captures, and
neutrino induced reactions with $\lambda_j$ being the rate per particle $j$.
The second term includes two particle 
reactions for particle densities $n_j,n_k$,  and mass density, $\rho$, with $N_A$ being  Avogadro's
number and $\delta_{jk}$ is the Kronecker delta. $\langle \sigma v \rangle_{j;k}$ is the nuclear cross-section convolved by the velocity of the particles $j$ and $k$ in the plasma.
The  $N^i_j$ and
$N^i_{j,k}$ are the number of particles $j,k$ created or destroyed in the
process. 
$O(3)$ includes terms due to multi-particle reactions.  
Here, we use 218 isotopes,
but the solver has been used for up to 2438 isotopes. For temperatures
larger  than $6.5 \times 10^9 K$, we assume nuclear statistical
equilibrium mediated by strong and photon induced reactions, with only terms due to weak reactions 
remaining in the rate equations. For $\langle \sigma v
\rangle_{i;k}$, Boltzman distribution is assumed for the velocity $v$ of the
isotopes in the plasma. The individual rates between isotopes are fitted by
expansions in $\rho$ and $T$  with factors tabulated in the reaction-network
library REACLIB and modified weak reactions. Updated cross-sections
are as published in \citet{Cy10}. 
For
more details, see e.g.  \cite{T94a,Langanke04}.  

\noindent{\bf Low density equation of state, opacities and source functions:}
The data for the atomic levels and line transitions are taken from the
compilation of Kurucz \citep{kur95}, \cite{seaton2005}, and \cite{linedat},
supplemented by additional forbidden lines from \cite{tiara15}. 
For the atomic
models, we reduce the number of energy levels by use of 
level-merging/super-levels, with the assumption that
coupling between the merged levels being in thermodynamical equilibrium.
This implies
full frequency redistribution of individual transitions (see below). For the
radiation transport,  
Voigt functions \citep{Voigt98} are assumed for the individual line profiles in comoving frame.
For Monte Carlo transport, we assume the narrow line limit
\citep{sobolev57,1971JQSRT..11.1365A,castor04} and take into account the
probability for absorption along the way by lines of higher frequency, similar
to the formulation by  \citet{karp77} and the narrow line-limit
\citep{sobolev57,1992ApJ...387..248H,2003ASPC..288..185H}.  The module for molecular
kinetics use rates from \citet{P89,S88,L90,sharp70,1995A&A...297..135H,gerardy07,Rho2020}.  
In addition, the time-dependent formation of dust for carbonates,
silicates and iron-crystals has been implemented as a kinetic gas
theory based on the codes for dust formation
\citep{Dominik1992,Dominik1997,Dominik2009}. 

{
We solve the full set of statistical equations to determine the population
density of atomic states, $n_i$, where $i$ is shorthand for the excitation state $i$ out of $js$ excitation states for the ionization state $k$ of element $el$.
 The time dependent level populations are given by the $el \times k
 \times j$ equations,
\begin{equation}
{\partial n_i \over \partial t} + \nabla (n_{i} {\bf v}) =
\sum_{i\ne j}
 (n_{j} P_{ji} - n_{i} P_{ij}) + n_{k'} (R_{k'i} + C_{k'i}) - n_{i} (R_{ik'} +
C_{ik'}).
\end{equation}
Here the rate, $P_{ij}$, is the sum of radiative, $R_{ij}$, and collisional,
$C_{ij}$, processes;  $k'$ stands for
all transitions from all bound levels of higher ionization states.
}

The non-thermal
excitation by hard $\gamma$-rays and electrons are added into the radiative
rates $R$ with fractions according to the individual level densities.  Time
scales in $\delta n_i/\delta t$ are dominated by the slow bound-free transitions
and, therefore, their time-dependence is taken into account only. Thus, the
time-dependent solution can be obtained from the stationary solution
$\tilde{n_i}$ by solving an inhomogeneous ODE analytically or as a simple system
of linear equations  \citep{h95} of the form
\begin{equation}
{d n_i \over d t} = - C_1 n_i + C_2  ~
{\rm with} ~
 C_1  = {\tilde{n}_k \over \tilde{n}_i} (R_{ki} + C_{ki}), ~~{\rm and}~~
 C_2  = n_k (R_{ki} + C_{ki})
\end{equation}.
The radiative rates between a lower and upper level $i$ and $j$,
respectively, are given by
\begin{equation}
  \tilde{R_{ij}} = 4 \pi \int
{\alpha_{ij} (\nu) \over h \nu} J_\nu d\nu + R(\text{non-thermal}), 
{\rm and }~
\tilde {R_{ji}} = 4 \pi  \int
{\alpha_{ij}(\nu) \over h \nu }
\biggl\lbrack {{2 h \nu^3 \over c^2} + J_\nu} e^{-{h \nu \over k T}}\biggl\rbrack  d\nu  .
\end{equation}
with $R_{ij}=(n_j/n_i)^*  \tilde{R_{ij}} $ and  $R(non-thermal)$ being the
energy input by hard-radiation and non-thermal electrons, $J_{\nu}$ is the first
momentum of the radiation field, $\alpha_{ij}$ is the cross-section, and $*$
denotes values for thermodynamical equilibrium.

The equation of particle and charge conservation can be written in the following form 
\begin{equation}
 \sum_{el} \sum_{k} \biggl\lbrack \sum_{levels} n_{el,k,j}
\biggr\rbrack =
n_{tot}
{\rm ~~and ~~}
\sum_{el} \sum_{k} Z_{k} \biggl\lbrack \sum_{levels} n_{el,k,j}
\biggr\rbrack =
n_e 
\end{equation}
where $el$, $k$ and $level$ are the sums of the element $el$, ionization stages $k$ for  element and
levels in the particular ion with Z$_{k}$ being the charge of the ion.
The electron and total densities are given by $n_e$ and $n_{tot}$,
respectively.
Complete redistribution over each individual line both in frequency and in angle
is assumed in the comoving frame. Complete redistribution also implies that the
relative populations within the sub-levels or the merged levels are described by
a Maxwell-Boltzmann distribution.

The formal total source function $S_\nu$, emissivity $\eta$, the opacity $\chi
$ and frequency redistribution function $\psi$ are related by 
\begin{equation}
S_\nu = {\eta_\nu \over \chi_\nu}  \psi_\nu
\end{equation}                      
with                            
\begin{equation}
\eta_\nu =                      
 {2 h \nu^3 \over c^2} \sum_{el}\sum_{k} \Biggl\lbrack  
\sum^I_{i = 1} \biggl\lbrack        
 n^*_i \alpha_{ik}(\nu) e^{-h\nu \over kT} +                        
 \sum^I_{j > i} (\alpha_{ij} (\nu) n_j {g_i \over g_j})\biggr\rbrack      + n_e n_{ion} \alpha_{ff} (\nu)
e^{-h \nu/k T}\Biggr\rbrack     
 + \Biggl \lbrack n_e \sigma_e + \sum_{\ell^* }                     
  \chi_{\ell^*}                        
 \Biggr\rbrack                      
 J_\nu                              
\end{equation}                       
and                                                     
\begin{equation}
\chi_{\nu} =                        
 \sum_{el} \sum_{k}\Biggl\lbrack   
\sum^I_{i = 1} \Bigl\lbrack \alpha_{ik} (\nu)                   
(n_i - n^*_i e^{-h\nu \over kT})    
+ \sum^I_{j > i} \alpha_{i j}       
(\nu) \Bigl(n_i - n_j {g_i \over g_j}\Bigr) \biggr\rbrack            
 + n_e n_{ion}                      
\alpha_{ff}(\nu) (1 -               
e^{-h \nu \over k T}) \Biggr\rbrack + \Biggl\lbrack                     
n_e \sigma_e + \sum_{\ell^*} \chi_{\ell^*}                 
\Biggr\rbrack                   
\end{equation}       
with $\alpha $ being the cross sections for bound-bound, bound-free and free-free transitions. $\chi_{\ell^*} $ are opacities of those weak scattering lines not treated in non-LTE.

 \noindent{\bf Line redistribution functions:} For the source function of low
 energy photons, complete redistribution is assumed in general for each
 individual bound-bound transition.  However, due to the large number of lines,
 they will overlap even in comoving frame due to their natural line width which
 results in a frequency redistribution in case of radiation transport in spherical geometry \footnote{as used in LC simulations}.  The redistribution
 function $\psi_\nu$ is evaluated numerically by weighting the neighboring
 frequencies with the overlap of a specific transition so that the frequency
 integral over $S_\nu$ is conserved
 \citep{1971JQSRT..11.1365A,1992ApJ...387..248H,1994ApJ...426..692R,h95}.\footnote{Note
 that the frequency coupling stabilize the iteration scheme.}

\noindent {\bf Coupling of radiation transport, statistical and hydro
equations:} We use the well established method of accelerated lambda iteration
(ALI, e.g.  \cite{cannon73,scharmer84,hillier90,h90,hubeny92}).  We employ
several concepts to improve the stability, and convergence rate/control
including the {\sl concept of leading elements}, the use of net rates, level
locking, reconstruction of global photon redistribution functions,
equivalent-2-level approach \citep{Athay92,avrloes88,h95},  and
predictor-corrector methods.

\noindent {\bf Radiation transport for low energy photons:} \hspace{1pt} For the
time-dependent transport, we use variable Eddington Tensor solvers, which are
implicit in time
\citep{MM84_book,stone1992,2003ASPC..288..185H,Castor07_book,castor09,2009AIPC.1171..161H}\footnote{Some
new additional VET modules are in the verification phase}, and stationary
solutions of the transport for the closures.

For spherical geometry, we solve the following equations
\begin{equation}
{1\over r^2} {\partial (r^2 H_{\nu}) \over \partial r} -
{\nu_o v(r) \over c r } \biggl\lbrack {\partial (1-f_{\nu}) J_{\nu}
\over \partial \nu} + \beta(r) {\partial f_{\nu} J_{\nu} \over
\partial \nu } \biggr\rbrack = \eta _{\nu} - \chi_{\nu} J_{\nu}
\end{equation}
and
\begin{equation}
{\partial f_{\nu} J_{\nu} \over \partial r } + { 3 f_\nu -1
\over r } J_{\nu} - {\nu_o v(r) \over c r }
\biggl\lbrack {\partial (1 - g_{\nu})
 H_{\nu} \over \partial \nu }
+ \beta (r) {\partial (g_{\nu} H_{\nu}) \over \partial \nu}
\biggr\rbrack =  - \chi_{\nu} H_{\nu} ,
\end{equation}
 where $\beta $ and the Eddington
factors are defined in the usual way
\begin{equation}
 \beta= {d~ln(v(r)) \over d~ln(r)},~~ f_{\nu}=K_{\nu}/J_{\nu}, {\rm and}~g_{\nu}=N_{\nu}/H_{\nu}.
\end{equation}
 $J_{\nu},H_{\nu},K_{\nu}$ and $N_{\nu}$ are the first four moments of the intensity. The 4th
 moment is needed as a closure  relation of the
system. 
The Eddington factors $f_\nu$ and $g_\nu$ are obtained from solutions for the
stationary case by integration along rays in a Rybicki-like scheme for
non-relativistic and relativistic velocity fields including advection terms
(\citep{mkh75,mkh76a,mkh76b}, MKH methods).  Note that the Eddington factors
includes the frequency redistribution function $\psi$.  Time independent Eddington
factors are assumed during each time step.

For the 3D case, the variable Eddington-Tensor method is used which is accurate
to the order O(u/v) and we neglect acceleration terms
\citep{buchler79,buchler1983,stone92,2009AIPC.1171..161H,Hubeny2014_book}.  The system of
equations is given by 
\begin{equation} {1\over c} {\delta F_\nu \over \delta
    t}+{1\over c} \nabla (u F_\nu + c \nabla (T_\nu E_\nu))=-k_\nu F_\nu
\end{equation}

\begin{equation}
\rho {D(E_\nu /\rho) \over D t} + E_\nu  T_\nu \nabla u - \nabla [ {c T_\nu \over \kappa_nu \rho q_\nu } \nabla (q_\nu E_\nu) ] = 4 \pi j_\nu - \kappa_\nu \rho c E_\nu 
\end{equation}

with the tensor T and, following Auer (1971), the variable  $q_{\nu} $ as defined by:

\begin{align}
\nabla{T_\nu \nabla log~ q_\nu } = \nabla (\nabla T_\nu )
\\
{\partial ln (r^2 q_{\nu}) \over \partial r } 
= { 3 f_{\nu} -1 \over f_{\nu} r}
\end{align}
and
\begin{equation}
    E_\nu (\vec{x}) = 1/c \int \vec{I_\nu} (\vec{x},\vec{\Omega}) d\vec{\Omega}, ~~
\vec{F}_\nu (\vec{x}) = 1/c \int \vec{ I_\nu}  (\vec{x},\vec{\Omega}) \vec{\Omega} \vec{d\Omega}, ~{\rm and}~~
 \underline{P}_\nu (\vec{x}) = 1/c \int \vec{I_\nu} (\vec{x},\vec{\Omega})  \vec{\Omega} \otimes \vec{\Omega} d\vec{\Omega}.
 \end{equation}
 with the integrals over $\nu$ denote the corresponding radiative components in
 the  hydro-equations. For spherical geometry, the pressure is taken as the
 diagonal elements of $\underline{P}$. For 3D, the advection frequency
 redistribution terms enter via the {Monte Carlo (MC)} solution. For some tests of 3D vs.
 spherical solutions, see \citet{h02a}.  Again, we use stationary transport for
 the closure relation and tested the result against the spherical transport
 module. {MC} methods are used
 including advection and aberration for the closure relation of low energy
 photons for flux and polarization spectra, and similar to hard $\gamma$ and
 positron transport \citep{h91,hkm92,h95pol,penney14}. 

\noindent{\bf Gamma-ray and positron  transport:} The $\gamma$-ray transport is
computed in  multi-dimensions using a  Monte Carlo method
\citep{hkm92,hoeflich2003hydra} including relativistic effects and a consistent
treatment of continuum and line opacities. Interactions are calculated in the
local rest frame by  transformation between observer and comoving frame.
Photon and positron packages may persist until the next hydro timestep.  If the photon/positron travel time exceeds the timestep, they're fed to the next timestep.  This persists until they are scattered down to low energies, X-rays, or escape the computational domain.  Each package keeps a time-tracer that monitors its travel time.

The interaction processes allowed  are:  Compton
scattering according to the full angle-dependent  Klein-Nishina formula,
pair-production, and bound-free transitions \citep{as88}, and the NIST XCOM data
base \citep{Berger}.  

\subsection{Positron Creation and Cross Sections:} The primary source of
positrons is the $\beta ^+$ channel of the $^{56}Co \rightarrow ^{56}Fe^* $
which accounts for about $18\%$ of all $^{56}Co$ decays.  About $1.4 MeV$ of the
total excitation energy of $^{56}Fe^*$ is available for the positrons with an
energy spectrum  \citep{SergeNuke} given by
\begin{equation}\label{eq:trans_spectrum}
    N(E)=Cp^2(E_o-E)^2(2\pi\eta(1-\exp(-2\pi\eta))^{-1}) \end{equation}  where
    $C$ is a scale factor and $\eta$ is the charge of the nucleus times $\hbar$
    over the velocity of the electron.  The mean energy of the spectrum is .44
    MeV.  

 For the Monte Carlo transport, we take into account three processes for the interaction: scattering on electrons, scattering on nuclei in a plasma, and annihilation 
 $e^+e^- \rightarrow 2\gamma$  or $e^+e^- \rightarrow 3\gamma$ for para- and ortho-positronium, respectively.

The annihilation  cross-section \citep{astroformula} is given by 
\begin{equation}\label{eq:electronsigma}
\sigma = \frac{\pi r_o}{\gamma+1}[\frac{\gamma^2+4\gamma+1}{\gamma^2-1}ln(\gamma+\sqrt{\gamma^2-1})-\frac{\gamma+3}{\sqrt{\gamma^2-1}}]
\end{equation} with $r_o=e^2/mc^2$ and $\gamma$ is the Lorentz \myeditt{factor}. 

Following  \citet{lingenfelter93} and \citet{gouldsigma}, the energy loss by interaction with  charged particles is given by
\begin{equation}\label{eq:bloch}
\frac{dE}{dx} = \frac{-4\pi r_o^2m_ec^2q}{AM_n\beta^2}(qln(\frac{\sqrt{\gamma-1}\gamma \beta m_ec^2}{b_{max}})\Pi (\gamma)
\end{equation}
$\Pi(\gamma)$ is the relativistic correction given by 
\begin{equation}
\Pi (\gamma) =
\frac{\beta^2}{12}[\frac{23}{2}+\frac{7}{\gamma+1}+\frac{5}{(\gamma+1)^2}+\frac{2}{(\gamma+1)^3}].
\end{equation}
$q$ is the relative charge of the particles in the media. For electron interactions, $q$ is 
the atomic number $Z$ for atom. For plasma scattering, $q$ is the ionization fraction.
$b_{max}$ is the maximum impact parameter.  It is the maximum amount of energy the positron can lose in
 one interaction.  For electron scattering, is it the ionization potential.  
For the case of ions in the plasma, the maximum impact parameter is $\hbar \omega$, which is the plasma frequency as an 
impact with an ion sets up a disturbance in the plasma whose energy is proportional to the frequency.
Upon undergoing either type of interaction, the particles  are emitted isotropically in the comoving frame. 

The positron transport \citep{penney14} is solved via a Monte Carlo method very similar to our photon transport but the integration is along a spiral path imposed by the local Lorentz force.

\vfill\eject

\end{appendix}

\end{document}